\documentclass[preprint2]{aastex62}
\graphicspath{{./}{figures/}}

\usepackage{amsmath}
\usepackage{graphicx}
\usepackage[T1]{fontenc}

\usepackage{txfonts} 
\usepackage[figure,figure*]{hypcap} 
\usepackage{chngcntr}

\begin{document}

\title{The breakBRD Breakdown: Using IllustrisTNG to Track the Quenching of an Observationally-Motivated Sample of Centrally Star-Forming Galaxies}

\shorttitle{The breakBRD Breakdown}
\shortauthors{Kopenhafer et al.}

\author{Claire~Kopenhafer}
\altaffiliation{kopenhaf@msu.edu}
\affil{Department of Physics and Astronomy and Department of Computational Science, Mathematics, and Engineering,\\ Michigan State University, 567 Wilson Rd, East Lansing MI 48823, USA}
\author{Tjitske~K.~Starkenburg}
\affil{Flatiron Institute, 162 5th Avenue, New York NY 10010, USA}
\affil{Center for Interdisciplinary Exploration and Research in Astrophysics (CIERA) and\\ Department of Physics and Astronomy, Northwestern University, 1800 Sherman Ave, Evanston IL 60201, USA}
\author{Stephanie~Tonnesen}
\affil{Flatiron Institute, 162 5th Avenue, New York NY 10010, USA}
\author{Sarah~Tuttle}
\affil{University of Washington, Seattle, 
3910 15th Ave NE, Room C319, 
Seattle, WA, 98195-0002, USA}

\begin{abstract}
The observed breakBRD (``break bulges in red disks") galaxies are a nearby sample of face-on disk galaxies with particularly centrally concentrated star formation: they have red disks but recent star formation in their centers as measured by the D$_n$4000 spectral index \citep{tuttleBreakBRDGalaxiesGlobal2020}. In this paper, we search for breakBRD analogues in the IllustrisTNG simulation and describe their history and future. We find that a small fraction (${\sim}4\%$ at $z=0$; ${\sim}1\%$ at $z=0.5$) of galaxies fulfill the breakBRD criteria, in agreement with observations.  In comparison with the mass-weighted parent IllustrisTNG sample, these galaxies tend to consist of a higher fraction of satellite and splashback galaxies.  However, the central, non-splashback breakBRD galaxies show similar environments, black hole masses, and merger rates, indicating that there is not a single formation trigger for inner star formation and outer quenching.  We determine that breakBRD analogue galaxies as a whole are in the process of quenching.
The breakBRD state---with its highly centrally concentrated star formation---is uncommon in the history of either currently quiescent or star-forming galaxies; however, approximately 10\% of $10^{10} < M_\ast/M_{\odot} < 10^{11}$ quiescent galaxies at $z=0$ have experienced SFR concentrations comparable to those of the breakBRDs in their past. Additionally, the breakBRD state is short-lived, lasting a few hundred Myr up to ${\sim}2$ Gyr. The observed breakBRD galaxies may therefore be a unique sample of outside-in quenching galaxies.
\end{abstract}

\section{Introduction}\label{sec:intro}

In the standard picture of galaxy formation, galaxies form in an ``inside-out'' sense: high densities and rapid gas cooling cause an early formation of the inner regions, while the outer parts form later due to lower densities and slower gas accretion and cooling
\citep[e.g.][]{Larson1976, MatteucciFrancois1989, Burkert1992, Chiappini1997, bosch1998, kepner1999}.

By far, most galaxies in the Local Universe seem to follow this general picture for their formation and growth, which can be traced through metallicity gradients and/or stellar age gradients and trends of scale radii with stellar age (\citealp[e.g.][]{PagelEdmunds1981, Shaver1983, Williams2009, Vila-Costas1992, Dale2016}; \citealp[see][for a review]{Maiolino2019}), even when taking into account processes such as stellar radial orbit migration \citep{Magrini2016,Frankel2019}. 
High resolution and spatially resolved observations largely corroborate the general ``inside-out'' picture  (\citealp[e.g.][]{Sanchez2014, Belfiore2017, LopezFernandez2018, SanchezMenguiano2018, Poetrodjojo2018}; \citealp[reviewed by][]{Sanchez2019}), but also reinforce a number of caveats. For example, outside-in growth has been observed in dwarf galaxies \citep{Gallart2008, Zhang2012, Perez2013, Pan2015, Ibarra-Medel2016, Wang2019}, and galaxy interactions may affect metallicity gradients through dry mergers as well as through triggering gas flows \citep{Mehlert2003, Sanchez-Blazquez2007, Queyrel2012}.

In more detail, the distribution of star formation within galaxies may be related to the global star formation rate.  For example, \citet{Morselli2019} found that star formation is centrally enhanced in galaxies near the star-forming sequence and centrally suppressed below it.  In agreement with this work, several authors have found that galaxies not only form inside-out but also quench from the inside-out  \citep{Ellison2018, Nelson2016, Li2015, Rowlands2018, Spilker2019}. Using the MaNGA survey \citep{Bundy2015,Yan2016}, \citet{Lin2019} find that the fraction of galaxies quenching inside-out grows with mass \citep[see also][]{belfioreSDSSIVMaNGA2018}.  Similar central sSFR suppression in galaxies below the star forming main sequence was found for both the CALIFA \citep{Sanchez2012} and SAMI \citep{Croom2012, Bryant2015} surveys \citep{Gonzalez-Delgado2016, Medling2018}. On the other hand, post-starburst galaxies, and galaxies with post-starburst or recently quenching regions, show large diversity  and irregularity in the distribution of their star-forming and quiescent regions \citep{Rowlands2018, Chen2019, Quai2019}.

The presence of a bulge may also affect the global star formation rate and the distribution of star formation within a galaxy (\citealp{Genzel2014,Mendez-Abreu2019, Martig2009, Gensior2020}, although see also \citealp{ Martig2013,Kretschmer2020,Su2019}).
Observations indicate that a large bulge component is often correlated with galaxy quenching, but that it is unlikely to cause quenching on its own \citep[e.g.][]{Bundy2010, Fang2013, Bell2012, Whitaker2015, Bait2017, Omand2014, Bluck2014, Gonzalez-Delgado2016}, and \citet{Lang2014} and \citet{McPartland2019} argue that the growth of the bulge precedes the global shutdown of star formation. 
However, when we focus on the disk, \citet{Abramson2014} have argued that excluding the bulge leads to a constant sSFR for the star forming disk component of galaxies.  In agreement, \citet{Medling2018} find that bulges have little effect on the star formation in the disks of even early-type galaxies. 

There are, of course, many factors that can affect star formation and its distribution within a galaxy.
Feedback from Active Galactic Nuclei (AGN) may decrease or even quench the SFR from the center outwards. It could play a critical role in the regulation of star formation \citep{Prasad2020, Voit2020} and is often argued to be required to keep galaxies quenched \citep[e.g.][]{Su2019}. AGN feedback is also related to bulge mass, illustrated in the M$_{BH}$-M$_{\rm bulge}$ relation (\citealp[e.g.][and references therein]{KormendyHo2013, HeckmanBest2014}; \citealp[but see also][]{Martin2018,Ding2020}). In large-scale cosmological simulations, the tuning of AGN feedback makes it the dominant feedback mechanism at low redshift: it decreases the SFR of highly star-forming objects \citep[e.g.][]{Katsianis2017, Dave2019} and is associated with quenching galaxies \citep[e.g.][]{Weinberger2018}. 

Major as well as minor mergers can bring in additional gas and increase star formation \citep{MihosHernquist1994, Cox2008, Kaviraj2014, Willett2015, Hani2020} and interactions can redistribute angular momentum and funnel gas toward the central regions \citep[e.g.][]{MihosHernquist1994, HernquistMihos1995, Naab2014, Lagos2018, Blumenthal2018}.  These processes may also be related to the possible correlation between mergers and AGN activity \citep{Ellison2011, Ellison2019, Goulding2018, McAlpine2020}.

Smaller  star-forming galaxies in dense regions tend to be influenced strongly by their  environment.  Satellite galaxy evolution can be driven by processes  such  as  tidal  gas stripping,  ram pressure stripping,  starvation,  or strangulation \citep{BoselliGavazzi2006}.  Indeed, some work has argued that ram pressure stripping can quench galaxy outskirts while enhancing central star formation \citep{Fujita1999, Koopmann2004, TonnesenBryan2012}. \citet{ByrdValtonen1990} use simulations to argue that the tidal field of clusters could drive cloud collisions and subsequent star formation. 

Recently, a population of observed galaxies was introduced by \citet[hereafter TT20]{tuttleBreakBRDGalaxiesGlobal2020}, called the ``Breaking Bulges in Red Disk" galaxies, or breakBRDs. These galaxies were selected to have had recent central star formation (using D$_n$4000) and red disks (using $g - r$ color). The parent sample consisted of a bulge/disk decomposed sample \citep{LacknerGunn2012} based on the galaxy sample and spectral information from the NYU Value Added Galaxy Catalog \citep[VAGC, ][]{Blanton2005} from the the Sloan Digital Sky Survey (SDSS) DR7 \citep{Abazajian2009}, while using images with improved sky subtraction from SDSS DR8 \citep{Aihara2011}. 

BreakBRDs appear transitional in the optical bands, while presenting as star-forming in the UV and IR bands.  Most of this sample has stellar masses above 10$^{10}$ M$_{\odot}$, so on the basis of mass they should present "inside-out" star formation.  In addition, TT20 found that \textit{both} the sSFR and the bulge-to-total (B/T) mass ratio in breakBRD galaxies tended to be larger than that in the star-forming parent sample galaxies (sSFR $>$ 10$^{-10.9}\ \mathrm{yr^{-1}}$), indicating that the large B/T ratio is not driving a decrease in sSFR.  Finally, breakBRD galaxies are well-distributed across environmental density in a similar fashion to their parent sample, indicating that environmental processes are not driving their unusual star formation distribution. 

In this paper we use the large-scale cosmological hydrodynamic simulation IllustrisTNG100 \citep{Nelson2018, Springel2018, Pillepich2018, Marinacci2018, Naiman2018} to gain insight into this unusual galaxy sample.
We note that it is well-known that simulations cannot recreate the observed universe with perfect accuracy or completeness.  For example, the quenched fraction of low-mass satellites is poorly reproduced in simulations \citep{Donnari2020, Dave2017}, there are too few low-mass black holes at $z=0$ in simulations \citep{Habouzit2020}, and the green valley galaxies in many simulations have too-centrally concentrated star formation \citep{Starkenburg2019}. However, we can account for this by comparing galaxies within the simulation to each other, rather than directly to observations. This way, we can find breakBRD analogue galaxies within the simulation and determine what makes them unique.  This investigation draws upon the strengths of cosmological simulations: for all galaxies, we know the stellar, gas, and black hole masses, the star formation rates, and the environment with complete accuracy. Moreover, and particularly interesting for studying galaxy evolution, we can trace galaxies through time to determine their evolutionary histories and futures. 

The IllustrisTNG100 simulation agrees well compared to observations for a number of galaxy characteristics, among them the $z=0$ galaxy color distribution \citep[][]{Nelson2018, pillepichSimulatingGalaxyFormation2018}. Moreover, while \citet{Starkenburg2019} show that simulated green valley galaxies can be too centrally concentrated compared to observations, green valley galaxies in IllustrisTNG100 show more diversity in radial star formation rate profiles.

In this paper, we begin by describing our selection of parent and breakBRD analogue samples in Section \ref{sec:methods}.  In Section \ref{sec:z00} we compare the breakBRD analogues identified at $z=0$ with the $z=0$ parent sample. We consider evolutionary pathways that may encourage breakBRD galaxy formation (Section \ref{sec:properties_drivers}) as well as the breakBRD growth history (Section \ref{sec:history}).  We then introduce the breakBRD analogues at $z=0.5$ (Section \ref{sec:z05}), comparing them to the concurrent parent sample in Section \ref{sec:z05prop}, and focusing on the future evolution of these galaxies in Section \ref{sec:future}.  In Section \ref{sec:discussion} we discuss our findings and compare our breakBRD analogues directly to the observed breakBRD population. Finally, we summarize our conclusions in Section \ref{sec:conclusion}.

\section{Methods}\label{sec:methods}
\begin{table*}
\begin{center}
\caption{Number of galaxies in our TNG100 parent sample and its subsamples}\label{tab:pop}
\begin{tabular}{llrrrr}
\tablewidth{0pt}
\hline
\hline
Name & Selection Criteria & $z=0.0$ & $z=0.03$ & $z=0.1$ 
    & $z=0.5$ \\
\hline
\textit{Parent} & $10 < \log (M_\ast/\mathrm{M_\odot}) < 12$ and $R_{1/2} > 2$~kpc & 6092 (63\%) & 6029 (63\%) & 5857 (63\%)
    & 5057 (67\%) \\ 
\textit{D$_n$4000} & \textit{Parent} $\cap \, \mathrm{D}_n4000 < 1.4$ for $r<2\ \mathrm{kpc}$ & 2650 (70\%) & 3008 (68\%) & 2918 (68\%)
    & 3196 (69\%) \\ 
\textit{g-r} & \textit{Parent} $\cap\, g-r > 0.655$ for $r > 2\ \mathrm{kpc}$ & 2816 (50\%) & 2579 (49\%) & 2383 (49\%) 
    & 890 (49\%) \\ 
\textit{bBRDa} & \textit{D$_n$4000} $\cap$ \textit{g-r} & 235 (37\%) & 288 (28\%) & 247 (29\%) 
    & 72 (19\%) \\ 
\hline
\end{tabular}
\end{center}
NOTE. - Selection criteria described in Sections \ref{sec:parent} and \ref{sec:bbrda}. 
The ``\textit{bBRDa}'' designation is short for ``breakBRD analogues''. Numbers in parentheses are the percentage of that subsample that are central galaxies.
\end{table*}

The IllustrisTNG100 \citep[public data release: ][]{ nelsonIllustrisTNGSimulationsPublic2019}\footnote{www.tng-project.org} is part of a suite of simulations run using the AREPO moving mesh code \citep{Springel2010} with upgraded subgrid models compared to the Illustris simulation \citep{vogelsbergerIntroducingIllustrisProject2014, Genel2014}; in particular, the upgrades modified the black hole accretion and feedback model \citep{Weinberger2017}, galactic winds \citep{pillepichSimulatingGalaxyFormation2018}, and added magnetohydrodynamics \citep{Pakmor2011}. TNG100 has a volume of  110.7 Mpc$^3$ and a mass resolution of $7.5 \times 10^6 \mathrm{M}_{\odot}$ and $1.4 \times 10^6 \mathrm{M}_{\odot}$ for dark matter and baryonic elements, respectively.
The gravitational softening is $0.74$ kpc at $z=0$ for the collisionless particles and adaptive with a minimum of $0.18$ kpc at $z=0$ for the gas cells.  The gas cells have a minimum (median) radius of 14 pc (15.8 kpc), and star-forming gas cells have a mean radius of 355 pc.

We select our sample of breakBRD analogues (also referred to as the bBRDa sample) and their parent sample from the TNG100 snapshots at redshifts $0$, $0.03$, $0.1$, and $0.5$, which were chosen for their lookback times of 0.48, 1.3, and 5.2 Gyr respectively. These times allow us to assess how long the breakBRD state lasts and determine their future evolution. In this paper we primarily focus on redshifts $0$ and $0.5$, looking at the bBRDa properties and their histories and futures. The other redshifts are included to help better understand how long the breakBRD analogue state might last. 
The analogue selections were made with a combination of photometric and spectral criteria, as described in Section \ref{sec:bbrda}.

\subsection{Parent Sample}\label{sec:parent}
\begin{figure}
\centering
\includegraphics[width=\columnwidth]{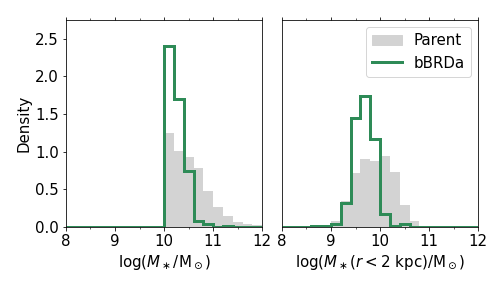}
\caption{\label{fig:masshistogram} Normalized histograms of stellar mass of the whole galaxy (left) and the inner 2~kpc (right) for our parent (solid grey) and breakBRD analogues (green) selection at $z=0.0$ (see Table \ref{tab:pop}). The break at $\log(M_\ast/\mathrm{M_\odot})=10$ is due to our selection criteria.}
\end{figure}

We define the ``parent'' sample of our analogues with two different criteria.  We first require that galaxy stellar mass must lie within $10^{10} < M_\ast < 10^{12}\ \mathrm{M_{\odot}}$. 
The lower mass limit was chosen out of concern for mass resolution: since our analysis requires that we directly measure properties in the central 2~kpc of our galaxies, we must be sure there is significant mass both in the entire galaxies and in the central regions.  Histograms of the total and central mass distributions of the parent sample at $z=0.0$ are shown in Figure \ref{fig:masshistogram}, and we see that even the lowest-mass central regions have more than several hundred particles.
We ignore galaxies with $M_\ast > 10^{12} \mathrm{\ M_\odot}$, as these have low numbers in TNG100, and can be assumed to be giant ellipticals.  Moreover, most ($\sim$80\%) of the galaxies in the \textit{observed} breakBRD sample have $M_\ast > 10^{10} \mathrm{\ M_{\odot}}$, and all have  $M_\ast < 10^{12} \mathrm{M}_{\odot}$ (TT20).

We also require galaxies in our parent sample to have $R_{1/2} > 2$~kpc, where $R_{1/2}$ is the stellar half mass radius. This removes galaxies which do not have a well-defined central region. 
The first row in Table \ref{tab:pop} lists the size of our parent samples at $z=0.0$, $z=0.03$, $z=0.1$, and $z=0.5$. The percentage of galaxies that are central galaxies is quoted in parentheses. 

\subsection{BreakBRD Analogue Sample}\label{sec:bbrda}

Our breakBRD analogues are selected from the parent sample 
using two additional criteria. These criteria are designed to mimic the selection criteria of the observational breakBRD sample, and select for galaxies with red disks and outskirts (and therefore little-to-no total star formation) but recent star formation in their centers. We accomplish this through an SDSS color cut of $g-r > 0.655$ for $r > 2$ kpc and a cut on the D$_n$4000 spectral measure of $<1.4$ for $r < 2$ kpc. The 4000 $\AA$ break is sensitive to the luminosity-weighted stellar age and is seen as an indirect measure of recent (${\lesssim}1$~Gyr) star formation \citep{hamilton1985, moustakas2006, brinchmannPhysicalPropertiesStarForming2004}.

The galactocentric radius $r$ is defined relative to the particle with the lowest gravitational potential in the galaxy. We use this instead of the galaxy center of mass, which may not reflect the galaxy's rotation center due to its sensitivity to structure at large radii \citep{genelGalacticAngularMomentum2015}. We use 2 kpc as the boundary between our inner and outer regions both to exceed the minimum spatial resolution of IllustrisTNG and to approximate the size of the SDSS spectral fiber at $0.003<z<0.05$, which is the redshift range of the breakBRD observational sample. These considerations are again why we restrict the parent to have $R_{1/2} > 2$~kpc.

In order to apply our selection criteria, we generate mock spectra of two radial bins for each parent galaxy, as well as for the whole galaxy. This requires star formation histories for both $r < 2$ kpc and $r > 2$ kpc. 
We combine the star formation history information of the star particles (in bins of 10 Myr) with the instantaneous SFR from the gas (added to the most recent bin), creating a full star formation history for both radial bins and for the entire galaxy. 

These star formation histories are then fed to the Flexible Stellar Population Synthesis (FSPS) code of \cite{conroyPropagationUncertaintiesStellar2009} \citep[updated in][]{conroyPropagationUncertaintiesStellar2010}, through the Python interface from \citet{danforeman-mackeyPythonfspsPythonBindings2014}, to generate mock spectra. For this generation we use the MIST isochrones \citep{choiMesaIsochronesStellar2016, dotterMESAIsochronesStellar2016, paxtonModulesExperimentsStellar2011, paxtonModulesExperimentsStellar2013, paxtonModulesExperimentsStellar2015} and the Miles spectral library \citep{sanchez-blazquezMediumresolutionIsaacNewton2006, falcon-barrosoUpdatedMILESStellar2011}. We use a \cite{chabrierGalacticStellarSubstellar2003} IMF and the \cite{charlotSimpleModelAbsorption2000} dust model with an effective absorption of $\tau = 1.0(\lambda/0.55\mathrm{\ \mu m})^{-0.7}$, which is reduced by a factor of 3 for stars older than $3 \times 10^7$ yr. This is the same dust model employed by \citet{torreySyntheticGalaxyImages2015} for stellar mocks of the original Illustris simulation. We emphasize that we only use the photometric colors and D$_n$4000 spectral index to select our breakBRD analogue (bBRDa) sample; omitting or changing the dust model in computing the spectra has only a small effect on our selection.

We generate separate spectra for the inner ($r < 2$ kpc) and outer ($r > 2$ kpc) regions, as well as for the whole galaxy for comparisons to observations (Section \ref{sec:obsv_compare}). We convolve the spectra with the SDSS bandpass functions to generate broadband magnitudes such as $g$ and $r$. We also calculated the D$_n$4000 measure for the spectra of the inner 2~kpc region following the narrow definition from \citet{baloghDifferentialGalaxyEvolution1999}.  Galaxies with $\mathrm{D}_n4000 < 1.4$ in the inner region have had star formation within the last $\sim1$ Gyr, and comprise the D$_n$4000 selection. This selection is composed of 2650 galaxies at $z=0.0$ and 3196 at $z=0.5$ (second row in Table \ref{tab:pop}). Galaxies with $g-r > 0.655$ outside 2~kpc comprise our $g-r$ selection (third row in Table \ref{tab:pop}). There are 2816 such galaxies at $z=0.0$, and 890 at $z=0.5$. 

The intersection of the D$_n$4000 and $g-r$ selections yields our breakBRD analogue galaxies, also referred to as the bBRDa galaxies. We refer to them as ``analogues'' because the observational sample in TT20 are the ``true'' breakBRDs. There are 235 such galaxies at $z=0.0$, 269 at $z=0.03$, 247 at $z=0.1$, and 72 at $z=0.5$ (final row in Table \ref{tab:pop}).

We will also occasionally reference the ``breakBRD state,'' which is the state in which a galaxy meets our D$_n$4000 and $g-r$ criteria. Since galaxies evolve over time, galaxies that were breakBRD analogues at one redshift may not qualify at another; we discuss this, and what we can infer from this about the length of the breakBRD state, in Section \ref{sec:future}.

\begin{figure}
\centering
\includegraphics[width=\columnwidth]{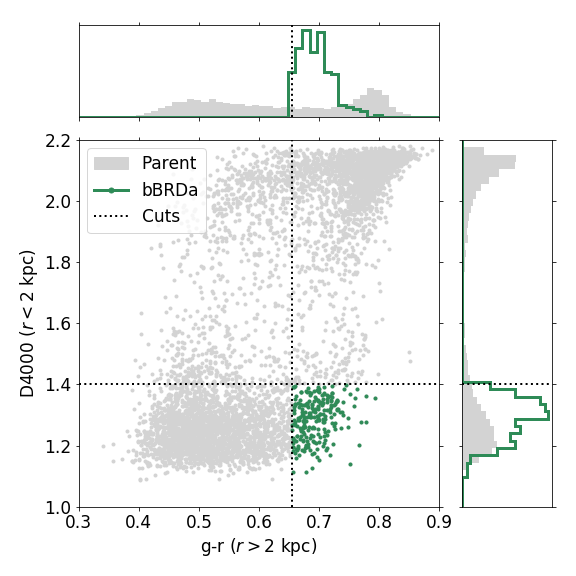}
\caption{\label{fig:D4000_color} D$_n$4000 in the inner 2~kpc vs $g - r$ color in the $r > 2\mathrm{\ kpc}$ disk for our selection at $z=0.0$. 
The parent sample is shown in grey, while the breakBRD analogues are green. Dashed black lines show the $\mathrm{D}_n4000<1.4$ and $g-r > 0.655$ selection cuts. Note that the histogram bins used for plotting straddle these cuts.}
\end{figure}

In Figure \ref{fig:D4000_color} we plot the D$_n$4000 values measured in the inner 2~kpc of parent galaxies at $z=0.0$ against their $g - r$ color in the outer region ($r > 2 \mathrm{kpc}$).  The breakBRD analogue subsample is separated out in green.  Each 1-dimensional distribution is shown in the histograms along the sides.  Clearly the breakBRD analogue sample is composed of an unusual subset of galaxies: most of the galaxies with central star formation have blue disks, and most of those with red disks have had little recent central star formation, but the bBRDa galaxies have star forming central regions with red disks.

The stellar mass distribution of our analogue sample is shown in Figure \ref{fig:masshistogram}, along with the stellar mass within 2~kpc. The total (and central) stellar mass of the bBRDa galaxies tends to be lower than that of the parent sample. Many of the quantities explored in our analysis are correlated with stellar mass; to that end, we weight the parent sample so that its mass distribution matches that of the bBRDa galaxies, as we discuss below. This weighting is applied from Figure \ref{fig:massSFR} onward.  We note that for all figures, the breakBRD galaxies are included in the parent samples. This does not affect our results.

\subsection{Sample Comparison}\label{sec:stats}

Many of the properties we examine in Sections \ref{sec:z00} and \ref{sec:z05} are dependent on both galaxy mass and galaxy classification as a central or satellite. We therefore separate centrals and satellites in our analysis for both the parent and breakBRD analogue samples. The fraction of central galaxies for each sample is given in parentheses in Table \ref{tab:pop}. 

To minimize the influence of stellar mass when examining differences between the parent and bBRDa galaxies (see Figure \ref{fig:masshistogram}), we apply a weighting to the parent central and satellites so that their stellar mass distributions match their respective analogue sample. 

\begin{table}
\begin{center}
\caption{Percentage of centrals that are splashbacks}\label{tab:splashbacks}
\begin{tabular}{lrrC}
\hline
\hline
Definition & bRRDa & Parent & $|z|$\\
\hline
After $z=0.1$ & 8\% & 4\% & 1.14\\
After $z=0.5$ & 23\% & 10\% & 2.26\\
\hline
\end{tabular}
\end{center}
NOTE. - Reported for the $z=0$ bBRDa and \textit{weighted} parent samples. Splashbacks are centrals that ``recently'' transitioned from being satellites, defined as either after $z=0.1$, or $z=0.5$ (see text). The fourth column gives the two-proportion z-test statistic comparing the two proportions. Proportions are significantly different at the $\alpha=0.05$ level if $|z|>1.96$, and at $\alpha=0.01$ if $|z|>2.58$.
\end{table}

As we are treating centrals and satellites separately, we also make a point to identify those galaxies that are splashbacks: galaxies that were considered to be satellites of more massive hosts at earlier times but have since escaped that halo. Splashback galaxies (also called backsplash galaxies or ejected satellites) are approximately related to a known caustic in the phase space structure of dark matter halos beyond the virial radius: the splashback radius \citep[see e.g. ][]{Adhikari2014, Diemer2017, Haggar2020, Diemer2020}\footnote{In IllustrisTNG the central/satellite definition depends on halos identified through the Friends-Of-Friends \citep[FOF, ][]{Davis1985} and SUBFIND \citep{Springel2001} algorithms, and are therefore not identical to central/satellite definitions based on halo radii determined through spherical overdensity criteria or phase space structure-motivated size definitions. \citet{more2011} show that corresponding spherical overdensities for FOF halos with linking length $b=0.2$ are correlated with on halo concentration and thus halo mass but are likely larger than the virial \citep{Bryan1998} overdensity. The resulting halo radii are therefore smaller than the virial radius, and also smaller than the splashback radius \citep[see e.g.][]{Diemer2020, Diemer2020b}. We conclude that there may be true splashback galaxies among our centrals.}, but we note that our definition can also include galaxies that interacted with the more massive host in a flyby encounter and were never gravitationally captured. Because, according to our definition, a splashback galaxy ``recently'' made the transition from satellite to central, 
it may exhibit properties similar to satellites \citep{Diemand2007, Knebe2011, Wetzel2014, Buck2019}. For $z=0.0$, we use both $z=0.1$ and $z=0.5$ as our cutoffs: if a galaxy has transitioned since the chosen redshift, it is counted as a splashback. 
We found that splashbacks only had an impact on the properties discussed in Sections \ref{sec:environ} and \ref{sec:merger}, as we discuss in those sections.  Table \ref{tab:splashbacks} contains the fractions of splashbacks in both the bBRDa and \textit{weighted} parent samples. 

Throughout the following sections we make use of three statistical tools to help quantify the differences and similarities between our various subsamples. First, in most figures, we have annotated the medians and median absolute deviations (MADs)\footnote{Defined as med$(|X_i - \Tilde{X}|)$ where $\Tilde{X}$ is the median of $\{X_i\}$}---robust statistical measures that are  resilient to outliers---for relevant distributions. 
As a reference, the $\mathrm{MAD}\approx 0.67\sigma$ for a Gaussian distribution; however, there is no strong indication that our data follow Gaussian distributions and so we keep our statistical evaluations distribution-free.

Second, we perform two-sample Kolmogorov-Smirnov (KS) tests to measure the ``distance'' between distributions and determine if it is statistically significant. Our tests take into account the weighting that we apply to the parent distributions to remove any dependencies on stellar mass. More sensitive tests, such as Anderson-Darling, exist; but, since we are trying to establish statistically meaningful differences rather than similarities, and the KS test functions best when differences are global, we consider the KS test to be the more conservative choice. 

Finally, there are a few instances in which we compare proportions instead of distributions. To determine if two proportions are statistically different from each other, we use a two-proportion z-test. This test functions by comparing the difference in proportions to zero. 

Both the KS test and the z-test have critical values for when to accept or reject the null hypothesis. For the two-sample KS test, the null hypothesis is that both empirical distributions are drawn from the same underlying distribution. For the two-proportion z-test, the null hypothesis is that the two proportions are equal. We use two common significance levels, $\alpha=0.05$ and $0.01$. A KS test result of $p<\alpha$ indicates the null hypothesis can be rejected. For the two-proportion z-test, these significance levels correspond to critical values of $|z|>1.96$ and $2.58$, respectively.

\section{breakBRD Analogue Galaxies in the Local Universe}
\label{sec:z00}
 
In this section we focus on the breakBRD analogues identified at $z=0$.  We first compare them to the $z=0$ parent sample, and then attempt to identify a formation mechanism for the breakBRD galaxies.

\subsection{Star Formation and Gas Properties}\label{sec:z00prop}

Here we explore the properties of the $z=0$ breakBRD analogue sample with respect to its parent population (see Sections \ref{sec:parent} and \ref{sec:bbrda} for their selection).

\subsubsection{Global Star Formation}\label{sec:ssfr}
\begin{figure}
\centering
\includegraphics[width=\columnwidth]{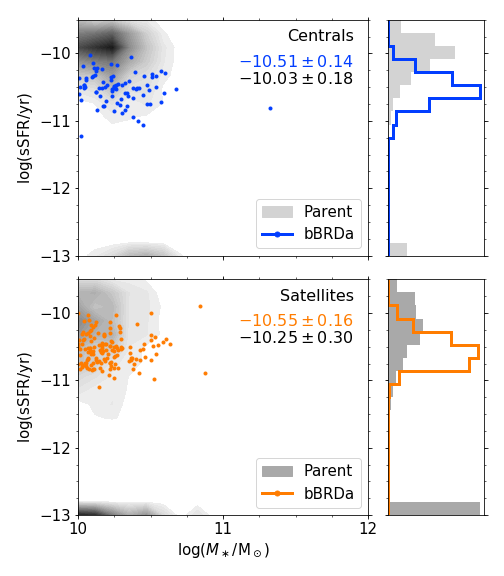}
\caption{\label{fig:massSFR}Total specific star formation rate vs total stellar mass for central galaxies (top) and satellites (bottom) in our $z=0.0$ selection. The weighted parent sample is represented by the grey contours. The weighting is described in Section \ref{sec:z00prop}. There is a floor imposed at $\log (\mathrm{sSFR/yr})=-13$, to which galaxies with no star formation were also set. The numbers in the upper right give the median and MAD when ignoring these galaxies.}
\end{figure}

We first consider where breakBRD analogues lie in the stellar mass-sSFR plane. Figure \ref{fig:massSFR} compares galaxy total stellar mass to total specific SFR (sSFR). 
Galaxies with no current star formation (SFR = 0) and those with $\log(\mathrm{sSFR/yr}) < -13$ were set to $\log(\mathrm{sSFR/yr}) = -13$. 
The top row shows central galaxies, and the bottom shows satellites\footnote{For all figures with contours, the background (white) contour level contains, on average, 5\% of parent galaxies. The percentage is at most 11\%.}. We remind the reader that in each case, as discussed in Section \ref{sec:stats}, the (central or satellite) parent sample has been weighted to match the total stellar mass distribution of the bBRDa centrals and satellites, respectively.

To consider this quantitatively, we calculate the median and median absolute deviations (MAD) of these distributions, taking into account the weighting, which are annotated on Figure \ref{fig:massSFR}. When calculating these values we ignore galaxies that were set to $\log(\mathrm{sSFR/yr}) = -13$. All of the bBRDa galaxies have $\log(\mathrm{sSFR/yr}) > -13$ and so have not been modified.

The two subsets of bBRDa galaxies cover roughly the same sSFR range, and tend to be lower in sSFR than each of their respective parents. Though the centrals' distributions overlap slightly, the median + MAD of the parent and breakBRD centrals do not, indicating each subsample is well clustered around their respective medians. 
The two bBRDa medians lie almost on top of each other and their MADs are of similar magnitude, in contrast to the two parent subsamples. The low sSFR of both breakBRD samples may be a signature of our selection criteria. 

This shows that the two bBRDa populations are more similar to each other in sSFR than they are to their respective parent populations, highlighting that in particular central breakBRD analogues are different from the general central  star-forming galaxy population.

\subsubsection{The Concentration of Stars and Gas}\label{sec:conc}
\begin{figure*}
\centering
\includegraphics[width=\textwidth]{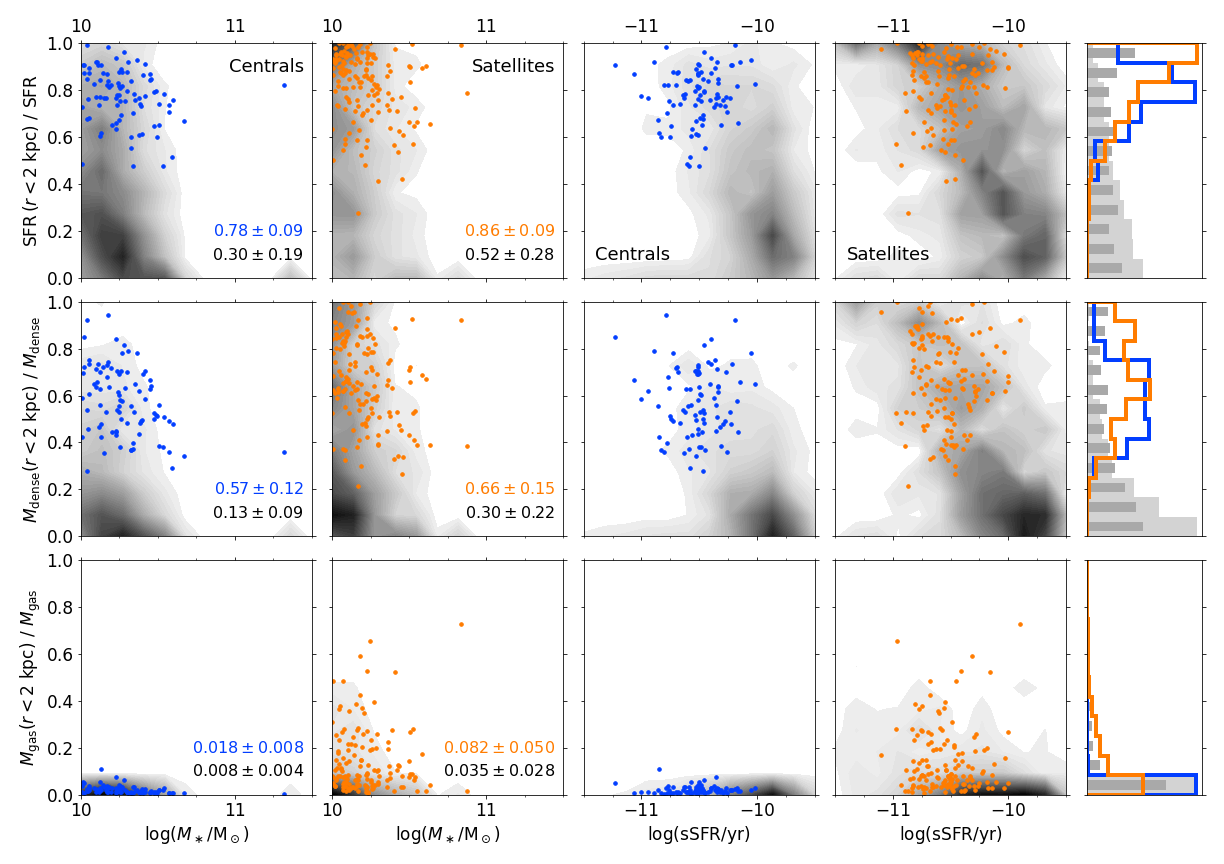}
\caption{\label{fig:multipanel}Concentration of star formation rate (top), mass of ``dense'' gas (middle), and total gas mass (bottom) within the inner 2~kpc of galaxies with nonzero SFR in our $z=0.0$ selection. ``Dense'' gas is gas above the density threshold for star formation in IllustrisTNG \citep{pillepichSimulatingGalaxyFormation2018}. These concentrations are plotted against total stellar mass (left two columns) and total specific star formation rate (next two columns) with normalized histograms (far right column). Both the weighted parent (contours) and breakBRD analogue samples (points) are split into central galaxies (blue) and satellites (orange). Median and MAD concentrations are labeled. The histogram colors follow Figure \ref{fig:massSFR}: filled for parent centrals (light grey) and satellites (dark grey bars), and lines for breakBRD analogue centrals (blue) and satellites (orange).}
\end{figure*}
\begin{figure*}[t]
\centering
\includegraphics[width=0.7\textwidth]{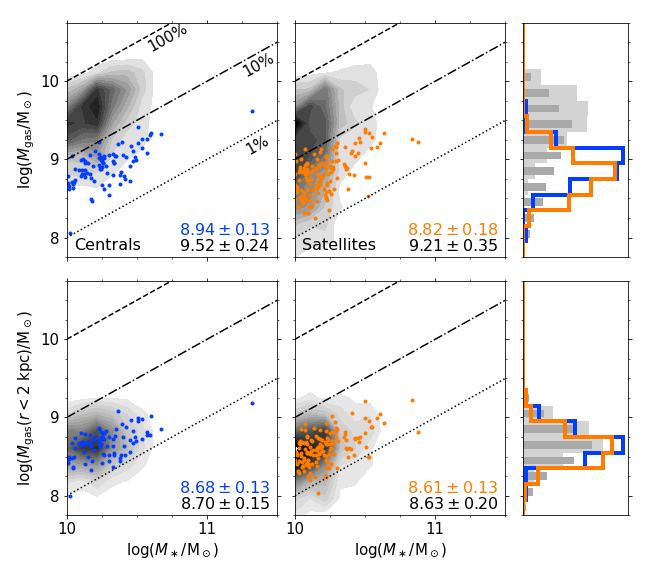}
\caption{\label{fig:gasmass}Total mass of dense gas (top) and mass of dense gas within the inner 2~kpc (bottom) vs total stellar mass for our selection at $z=0.0$. ``Dense'' gas is defined by the SF threshold in IllustrisTNG \citep{pillepichSimulatingGalaxyFormation2018}. Both the weighted parent (contours) and breakBRD analogue samples (points) are split into central galaxies (left) and satellites (right). Guides show where the gas mass is 100\% (dashed), 10\% (dot-dashed) and 1\% (dotted) of the stellar mass. Medians and MAD gas masses are labeled; they change by less than $0.2\%$ when splashbacks are removed. See Figure \ref{fig:multipanel} for a description of the normalized histograms to the far right.}
\end{figure*}

In Figure \ref{fig:multipanel} we examine the concentration of star formation and gas using the ratio of the inner 2~kpc of galaxies to their whole. From top to bottom, the three concentration ratios we examine are in star formation rate, ``dense'' gas mass, and total gas mass. Dense gas is that which is above 0.1 $\mathrm{cm^{-3}}$, the threshold for star formation in IllustrisTNG \citep{pillepichSimulatingGalaxyFormation2018}. Each ratio is plotted against both total stellar mass and total instantaneous sSFR, the two quantities from Figure \ref{fig:massSFR}. Once again, the parent and breakBRD analogues are broken into central and satellite galaxies. For some parent galaxies, these ratios are undefined: 1439 parent galaxies (563 centrals) have no SFR at all, 1385 (517 centrals) have no dense gas, and 620 (9 centrals) have no total gas. None of these galaxies belong to the breakBRD analogue subsample. These galaxies are naturally excluded from Figure \ref{fig:multipanel} and are ignored in the ensuing discussion. We apply the same weighting to the parent sample as used in Figure \ref{fig:massSFR}, and even though some galaxies are excluded, the mass distributions of the weighted parents remains similar to those of the bBRDa galaxies for all ratios plotted (the lowest $p$-value from two-sample KS tests is 0.80).

The combination of the $g-r$ photometric selection and D$_n$4000 spectral selection detailed in Section \ref{sec:bbrda} is expected to select galaxies with centrally concentrated star formation. The top row of Figure \ref{fig:multipanel} shows that we have indeed selected galaxies with this property. For both central and satellite galaxies, the concentration of SFR in the breakBRD analogue galaxies is clustered above 50\%. While some parent galaxies, especially satellites (dark bars), also have high SFR concentrations, their overall distribution is much flatter. Among the bBRDa galaxies, the satellites tend to be slightly more concentrated than the centrals, but their MADs are roughly the same and both distributions skew towards high concentrations. The parent satellites also have a higher median than the parent centrals, but again, both distributions are quite broad. We note that classification as a parent or a breakBRD has a stronger impact on SFR concentration than being a satellite or central, with the bBRDa galaxies consistently more concentrated. The two breakBRD subpopulations are also more similar to each other than the parent populations.

Given that star formation requires cold, dense gas, it's not surprising that in the middle row of Figure \ref{fig:multipanel} we see that dense gas is also more centrally concentrated in the bBRDa galaxies than in the parent sample. The dense gas concentration in the breakBRD analogues never falls below 20\%. The bBRDa populations have overlapping median + MAD ranges, though as with the SFR, satellites exhibit slightly higher concentrations.  The difference in concentration between the parent and bBRDa galaxies is once again more pronounced than the differences between centrals and satellites in either sample.  We also highlight that the dense gas concentrations of breakBRD satellites and centrals are more alike than those of the parent satellites and centrals.

Finally, the bottom row of Figure \ref{fig:multipanel} shows the degree to which \textit{all} gas in the galaxies is centrally concentrated. Both classifications of parent and breakBRD galaxies tend to have concentrations near 0. This trend is stronger for the centrals: 99.5\% of parent centrals have 5\% or less of their total gas in their inner 2~kpc, compared to 74.4\% of parent satellites. As satellites are expected to experience ram pressure stripping, this difference is not surprising.
For both satellites and centrals, the breakBRD analogues tend to have slightly higher concentrations than their corresponding parents, but these differences are not as strong as for the other quantities in Figure \ref{fig:multipanel}.
Because all of the populations skew towards zero concentration, we use a two-sample Kolmogorov-Smirnov (KS) test to further distinguish the parent and bBRDa galaxies. This KS test accounts for the weighting in the parent (see Section \ref{sec:stats}). Testing centrals against centrals and satellites against satellites yields $p$-values of order $10^{-13}$ or less, indicating the breakBRD analogues and parent samples are \textit{not} drawn from the same underlying distribution.

For all of the quantities shown in Figure \ref{fig:multipanel}, the breakBRD analogues always have much higher concentrations than the parent sample. We clearly see that for the concentrations of dense gas and SFR, the two breakBRD subpopulations are more similar than the two parent subsamples. These trends are unchanged when splashbacks are removed from the central subsamples.

With Figure \ref{fig:gasmass}, we re-examine the spatial distribution of dense gas in our galaxies in order to determine whether this gas is more concentrated in the center of galaxies because the central regions have additional gas, or because dense gas is missing from galaxy outskirts. 
As a function of stellar mass, we plot the galaxy's total dense gas mass in the top row, and the mass of dense gas within 2~kpc in the bottom.  
Central galaxies are on the left and satellite galaxies are on the right. Lines are included to guide the eye: they show where the gas mass is equivalent to 1\%, 10\%, and 100\% of the stellar mass. 
As with Figure \ref{fig:multipanel}, we exclude parent galaxies that do not have any dense gas (1385 total, 517 centrals), or lack dense gas in their inner 2~kpc (1610 total, 1199 centrals). 
Figure \ref{fig:gasmass} again uses the same weighting as Figure \ref{fig:massSFR}. The exclusion of the aforementioned galaxies does alter the mass distribution of the weighted parent sample, but KS test $p$-values remain acceptable ($p\sim0.8$ for the total mass distributions and $p\sim0.2-0.3$ for the inner gas).

The top row of Figure \ref{fig:gasmass} shows that 
both bBRDa subsamples are systematically lower than their respective parents in total dense gas mass, and are more similar to each other in terms of median and MAD than the two parent samples. Specifically, the central bBRDa galaxies are significantly different from the parent central distribution.
This result also holds when splashbacks are removed from the central samples (the medians and MADs change by less than $0.2\%$).

In contrast to the total mass of dense gas, all of our galaxy subpopulations have very similar distributions for the mass of dense gas in their inner 2~kpc (bottom row of Figure~\ref{fig:gasmass}). We clearly see that their medians and MADs all lie on top of each other, and unlike Figure \ref{fig:multipanel} and the total dense gas mass, the difference in central gas mass is more pronounced between satellites and centrals than between parents and breakBRDs.
This is born out when using a two-sample KS test: comparing satellites to satellites and centrals to centrals results in $p$ values above a threshold of 0.05, while comparing the two bBRDa (and parent) samples to each other yield $p$ below 0.01.  Removing splashbacks from the central samples only strengthens the similarity between those two distributions.
These results suggest that breakBRD galaxies exhibit an outer deficit of gas and not an central enhancement.

We note that Figure \ref{fig:gasmass} is qualitatively the same when we use \textit{all} of the bound gas instead of only the dense gas in the galaxies.  The total gas masses are low in the bBRDa galaxies while the central gas masses are similar in all subpopulations.

In summary, we find that the breakBRD analogues have more centrally-concentrated SFRs than the weighted parent samples. Dense, SF-eligible gas is also more centrally concentrated in the bBRDa galaxies. Being a breakBRD is a stronger predictor of concentration for these two quantities than being a central or satellite. 
When considering all of the dense gas, the two breakBRD subsamples are both lower than their respective parents, but all subpopulations have roughly the same mass of dense gas in their central 2~kpc. The breakBRD analogues therefore seem to be missing gas in their outer regions with respect to the parent galaxy sample.

\subsection{Possible Evolutionary Drivers}\label{sec:properties_drivers}

Thus far we have shown that the galaxies in the breakBRD analogue sample at z$=$0 lie somewhat below the star-forming main sequence (Figure \ref{fig:massSFR}), and their central concentration of star formation seems correlated with a lack of gas and star formation in their outer regions (Figures \ref{fig:multipanel} and \ref{fig:gasmass}).  Here we focus on three mechanisms that may drive galaxy evolution; namely, black hole feedback, mergers, and environment.  For each we determine whether breakBRD analogues differ from the parent population.   

\subsubsection{Black Hole Mass}\label{sec:bhs}
\begin{figure}
\centering
\includegraphics[width=\columnwidth]{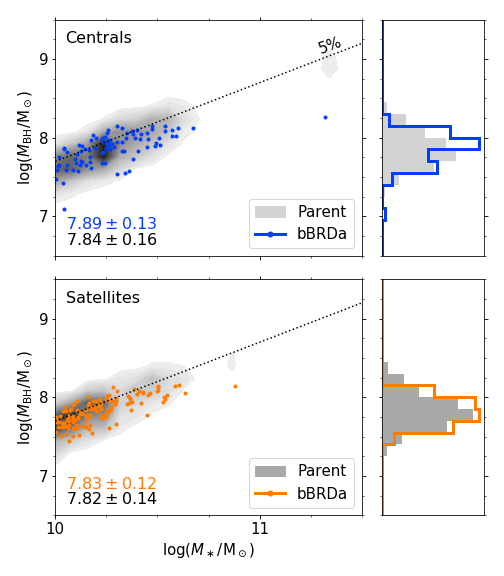}
\caption{\label{fig:BHmass}Black hole mass vs total stellar mass for central galaxies (top) and satellites (bottom) in our $z=0.0$ selection. The weighted parent sample is represented by the grey contours. 
A line where the black hole mass is 5\% of the total stellar mass is included to guide the eye. Medians and MADs for the black hole mass are labeled.}
\end{figure}

It has often been argued that there is a correlation and co-evolution of galaxies, or the central regions of galaxies, and their central supermassive black holes \citep[see e.g.][]{HeckmanBest2014}. Thus, the centrally-concentrated gas and star formation in breakBRD galaxies could also be reflected in their black hole mass. We might expect high black hole mass because high central gas density may enhance black hole accretion and possibly also AGN activity.  The causal link may be reversed, however, and strong AGN feedback from massive black holes may remove gas from a galaxy's outskirts.  Indeed, in IllustrisTNG AGN feedback is an important mechanism quenching star formation in galaxies \citep{Weinberger2018}.  

Therefore, we now look at the mass of the central black hole in breakBRD analogues as compared to the mass-weighted parent sample in Figure \ref{fig:BHmass}. This is plotted against total stellar mass. A line is included where the black hole mass is 5\% of the total stellar mass only to guide the eye. Some parent galaxies do not have black holes (their masses are stored as negative infinity). There are 121 such satellites, 10 of which are also breakBRD analogues. There is also one parent central without a black hole. These have been omitted from Figure \ref{fig:BHmass} and our analysis, and do not significantly impact the weighting of the parent sample.

The black hole mass distributions of the central and satellite breakBRD galaxies match well with their parent distributions. As seen in Figure \ref{fig:BHmass}, their medians and MADs are all very similar. This result remains when we instead calculate these statistics for $\log(M_\mathrm{BH})/\log(M_\ast)$, where the statistics are all $\sim0.77\pm0.01$.

Additionally, we have performed a similar comparison of the black hole accretion rates, both the present-day accretion rates as well as the maximum black hole accretion rates for each galaxy since $z < 0.03$ and since $z < 0.5$. Because AGN feedback is related to black hole accretion rates \citep{Weinberger2018}, if breakBRD galaxies had experienced strong AGN feedback, the maximum black hole accretion rate within their recent history would likely be high as well. Yet the median + MAD accretion rate ranges heavily overlap between the breakBRD analogues and their parents, whether using the rates at $z=0$ or finding the maximum recent accretion rate since $z=0.5$.  There is no evidence that the growth of the BHs in breakBRDs differs from that of the parent sample.

Using weighted two-sample KS tests to statistically assess the similarity of the distribution shape (Section \ref{sec:stats}), we cannot reject the null hypothesis that the parent and analogue satellites or centrals are drawn from the same underlying distributions ($p>0.1$ for both). Both of these test results support what is seen by eye in Figure \ref{fig:BHmass}, that the black hole mass distribution is very similar in the bBRDa and parent galaxy populations. This result is unaffected by the removal of splashbacks.
We thus find that breakBRDs sit on the same black hole mass -- stellar mass relation as the parent sample and do not have unusual black hole masses.

\subsubsection{Environment}\label{sec:environ}
\begin{figure}
\centering
\includegraphics[width=\columnwidth]{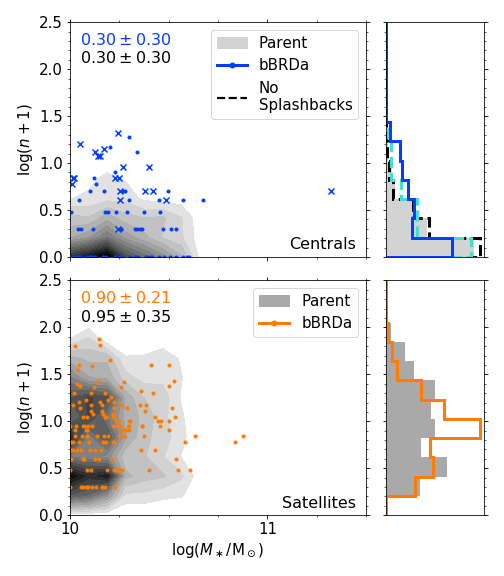}
\caption{\label{fig:environment}Number of galaxies with $M_\ast > 10^{10} \mathrm{\ M_\odot}$ within 2 Mpc vs total stellar mass, broken into centrals (left) and satellites (right), for our $z=0.0$ selection. The number $n$ of neighboring galaxies is plotted as $\log(n+1)$. The weighted parent sample is represented by the grey contours and the histograms are normalized. Dashed lines on the centrals' histogram are the distributions if galaxies that have been splashbacks since $z=0.5$ are removed. BreakBRD splashbacks are marked (crosses); these galaxies tend to live in the densest environments. Medians and MADs for $\log(n+1)$ are labeled for the full sample, including splashbacks. We note that the host galaxies of the satellites are not excluded from our environment measurement for either the parent or breakBRD samples.}
\end{figure}

As we have discussed in the introduction, satellite galaxies are more likely to experience environmental effects that may quench their outskirts by removing gas while enhancing star formation in the inner regions.  We might expect breakBRD galaxies to only reside in dense environments; although, as we discuss in Section \ref{sec:obsv_compare}, the observed sample shows no environmental influence.  Thus, in this section we consider several indicators for the environment of breakBRD analogues and the parent sample. 

First, we can simply compare the satellite fractions in the bBRDa and parent samples using Table \ref{tab:pop}.  The parent sample has a $\sim$37\% satellite fraction while the breakBRD analogue sample has a $\sim$63\% satellite fraction. This difference is statistically significant according to a two-proportion z-test ($|z|=4.49$; $|z| > 2.58$ is significant at $\alpha=0.01$). 
Given that there are multiple potential pathways for satellites to become centrally concentrated, the enhanced satellite fraction among breakBRDs is not a surprise; however, there remains a large fraction of central bBRDa galaxies.

In Section \ref{sec:stats} we introduced Table \ref{tab:splashbacks} and the fraction of splashback galaxies in the analogue and weighted parent samples. In both cases, the proportion of splashbacks in the bBRDa sample is double that in the parent sample; however, the statistical significance of this difference depends on the redshift used. The table includes the two-proportion z-test statistic, which tells us whether to reject the null hypothesis that the two proportions are statistically the same. At the $\alpha=0.05$ level, the breakBRD analogues have a significantly higher proportion of splashbacks since $z=0.5$. The proportions of splashbacks since redshift $z=0.1$ are generally low, which likely leads to the lack of significance under this definition. Thus only 77\% of central breakBRD analogues have not previously been satellites (since $z=0.5$) compared to 90\% of the parent.

Within our central and satellite galaxy selections we can also measure the environmental density.  In Figure \ref{fig:environment}, we look at the number of $M_\ast > 10^{10} \mathrm{M_\odot}$ galaxies within 2 Mpc as a measure of the density of surrounding galaxies. The number of neighboring galaxies that meet this criteria are plotted as $\log(n+1)$. 

First we note that the median and MAD values for the parent and bBRDa distributions are quite similar.  They are quoted in $log (n+1)$ space in the figure, so are equivalent to 1 galaxy for both central samples and 7 and 8 galaxies for the bBRDa and parent satellite samples, respectively.  There is significant overlap in the MADs.

When we remove either kind of splashback (from $z =$ 0.1 or 0.5; see Table \ref{tab:splashbacks}), the parent's median density drops from 1 galaxy with $M_\ast > 10^{10} \mathrm{M_\odot}$ within 2 Mpc to 0 galaxies, while the bBRDa median density remains at 1 galaxy.  We see that removing splashbacks decreases the width of the tail in the centrals' histogram at higher densities; this is shown with the dashed lines in the upper right of Figure \ref{fig:environment}. This is to be expected: since splashbacks are galaxies that were recently satellites, we should find them in denser environments. However, even with splashbacks removed, the MADs overlap for the parent and breakBRD central samples.

For a closer look at the environmental density distributions we also performed two-sample KS tests.  The breakBRDs' density distributions are significantly different from those of the parents ($p=0.011$ for the satellites, and the centrals have $p=7\times 10^{-5}$).  The $p$-value for the centrals increases as we remove galaxies that have been splashbacks since earlier times, reaching $p=0.29$ when removing those since $z=0.5$. The non-splashback central breakBRD galaxies thus live is similar environments to the non-splashback central parent sample.

While the two satellite distributions have an insignificant $p$-value, they cover roughly the same density range. Because the median values are close and the MADs overlap, even though the shapes of their distributions are different, we do not consider the satellite breakBRDs to be in meaningfully different environments than the parent sample satellites.

Additionally, we check whether these results change if we look at the distance and mass of the nearest more massive galaxy, or at the distance and mass of the nearest cluster (where we define cluster as a halo with M$_{DM} > 10^{14}\  \mathrm{M}_{\odot}$). We find that the environment of breakBRD analogues are similar using both of these measures. This similarity exists in both the shape of the distribution measured using the KS test, and in the close overlap of the medians and MADs.

Therefore, while the fraction of current or recent satellite galaxies is higher in the breakBRD than in the parent sample, the environment of galaxies within either the satellite or central samples is quite similar to the corresponding parent sample. This may indicate that a more subtle environmental effect is responsible for the evolution of breakBRDs and that this avenue is worthy of further investigation.

\subsubsection{Merger History}\label{sec:merger}
We counted the number of central and satellite galaxies that experienced at least one merger with mass ratio $M_{*,2}/M_{*,1} \ge 0.01$ since $z=0.5$. 
Mergers were classified by the stellar mass ratio of the galaxies involved:
$M_{*,2}/M_{*,1} \ge 1/3$ are defined as as ``major'' mergers while those from $1/10$--$1/3$ are minor \citep{Crain2015,Lagos2018,Katsianis2019}. We also include the $1/100$--$1/10$ range, which is composed of very minor mergers that could be considered cosmological accretion. We compare the fraction of galaxies that experience at least one of our defined events and determine statistical significance using two-proportion z-tests (see Section \ref{sec:stats}); note that the parent fractions include the mass weighting that has been applied throughout this paper. 

The frequency of major and minor mergers is not significantly different between the breakBRDs and their weighted parent samples. This is true even though \textit{none} of the breakBRD galaxies have experienced a major merger, because only $\sim$2\% of galaxies in the weighted central and satellite parent samples have. This lack of significance holds when we remove centrals that have been splashbacks since redshift 0.1 and 0.5.

According to our z-tests, the only significant difference is in the frequency of  
\textit{very} minor mergers (mass ratios of $1/100$--$1/10$) experienced by the centrals: 26\% of the parent centrals experience at least one accretion event, versus 10\% of breakBRD centrals. This is significant at $\alpha=0.01$, even when removing splashbacks since redshift 0.1, and is significant at $\alpha=0.05$ when removing splashbacks since redshift 0.5.

The decreased frequency of low mass accretion events can be interpreted as a sign of less cosmic accretion overall (including smooth, which this method poorly accounts for), as small satellites are often accreted onto more massive halos together with general accretion along a filament. This makes it unlikely that breakBRD analogue galaxies are generally formed by gas being brought in from outside the halo or funneled from the outskirts to the central region through tidal effects. Both central funneling and the addition of new gas are mechanisms that would increase the central concentration of gas and star formation rate as seen in Section \ref{sec:conc}. However, the dearth of recently experienced minor accretion events could point to indirect and subtle starvation-like environmental effects on the evolution of the central breakBRD galaxies.

\subsection{The History of breakBRDs}\label{sec:history}
\begin{figure*}
\centering
\includegraphics[width=\textwidth]{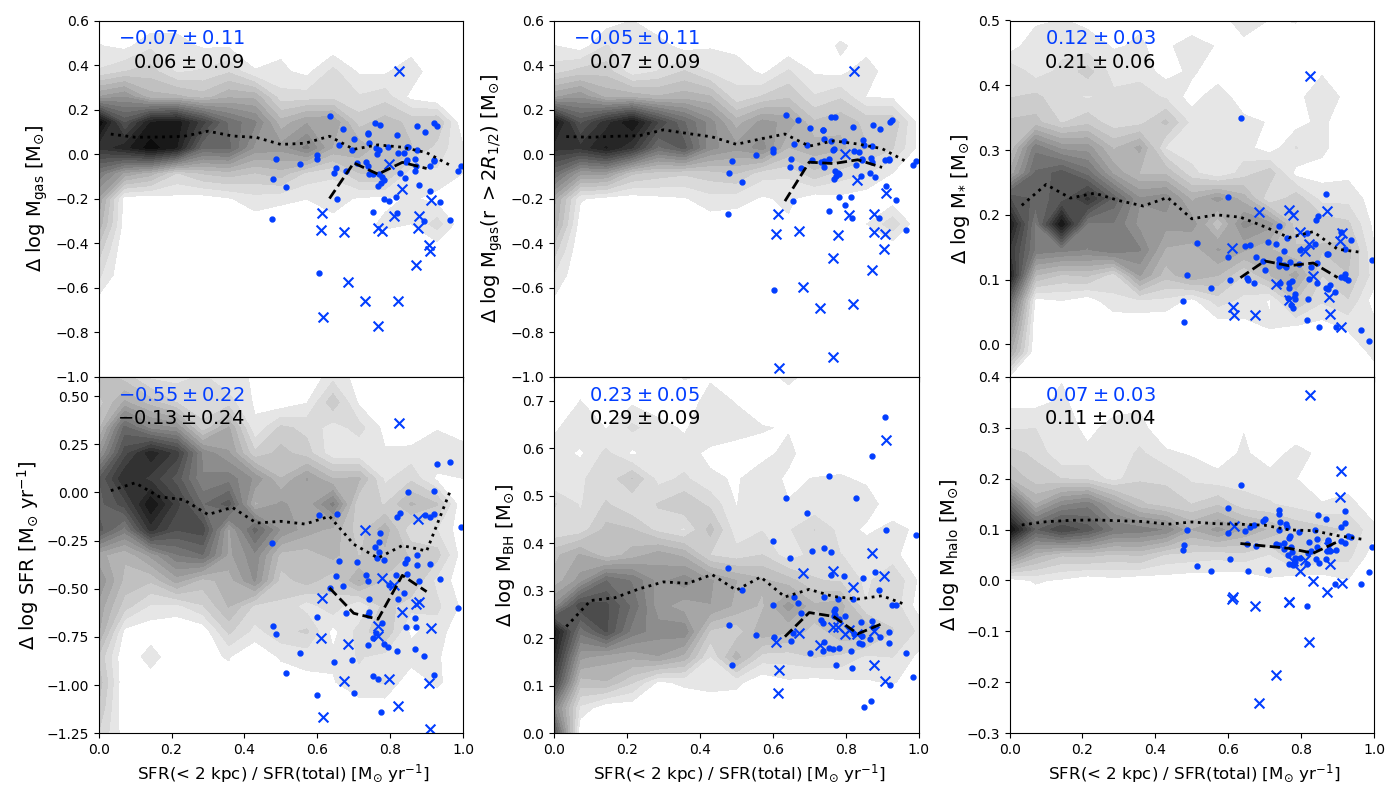}
\caption{Present day star formation rate concentration against relative growth (${\rm log} X(z=0.0) - {\rm log} X(z=0.5)$) of total gas mass (top left), star formation rate (bottom left), outer gas mass (top middle), supermassive black hole mass (bottom middle), total stellar mass (top right), and total halo mass (bottom right), since $z = 0.5$ for the parent central galaxies sample (grey contours) and our breakBRD-analogues central galaxies sample (blue scatter points). Running median trends are highlighted in black for both the parent sample (dotted) and our breakBRD-analogues (dashed), with the overall medians and MADs in the top left corner. The parent central sample is mass-weighted to its respective bBRDa central sample. Splashback breakBRD centrals are indicated (crosses).}
\label{fig:z0_history}
\end{figure*}

In an effort to understand why and how breakBRD analogue galaxies have their unique star-forming and color characteristics, we now trace the history of the $z=0.0$ sample. While the current properties of breakBRD galaxies give us some insight into how they differ from the parent sample, tracking their evolution allows us to understand these differences in more detail.
Moreover, as Table \ref{tab:overlap} shows that none of the $z=0.0$ breakBRDs qualify as breakBRD galaxies at $z=0.5$, the $z=0.0$ breakBRDs must have evolved into the breakBRD state somewhere in this window, and we may identify the physical drivers of this evolution by studying their history since $z=0.5$.

Figure \ref{fig:z0_history} shows the relative growth (the differences in log, or ratio, of the amounts at $z=0.5$ and at $z=0$)
of a number of galaxy properties since redshift $z=0.5$ for the redshift $z=0$ bBRDa and parent \emph{central} galaxies: the total gas mass, the gas mass outside the disk, the total stellar mass, the total star formation rate, the total black hole mass, and the total halo mass. This is compared against the SFR concentration at the present day. The parent sample is mass-weighted with respect to the mass distribution of the breakBRD galaxies at $z=0$. Overplotted lines indicate the rolling median trends for the parent (dotted) and breakBRD analogues (dashed) samples. We choose to show the relative growth with time against the $z=0$ SFR concentration because that is one of the most distinguishable features of the breakBRD analogue galaxies (as we have shown in Section \ref{sec:z00prop}). We have found that the bBRD satellites experience similar evolution as the parent satellite sample and therefore show the central galaxies only. 

While the total gas mass for the central parent sample has predominantly grown for non-concentrated galaxies, and is stable for galaxies with high star formation rate concentration, the central breakBRD analogues have almost all lost significant amounts of gas (up to $0.77$ dex, or 68\%; see the upper left panel of Figure \ref{fig:z0_history}).
The most dramatic gas loss in breakBRD analogue galaxies is for splashback galaxies, without which the breakBRD analogues lie closer to the parent sample with a median and MAD of $-0.04 \pm 0.07$.

We note that, in contrast to central galaxies, the majority of satellite galaxies have lost gas since $z=0.5$. This trend is stronger for more centrally concentrated satellites. This is to be expected if these satellites lose their gas through, for example, ram-pressure stripping.  

When we check the growth of the gas mass in the outer regions of the galaxies, we find that the mass in total gas that is lost is predominantly lost from the outskirts ($r > 2R_{1/2}$, see the upper middle panel of Figure \ref{fig:z0_history}). While not shown, we have checked that the gas evolution in the inner parts is similar to that of the parent sample (akin to the differences between the upper and lower panels of Figure~\ref{fig:gasmass}).

Additionally, we check whether the amount of gas that is lost is converted into stars, as opposed to being removed from or never accreted by the galaxy. About half of the central bBRDa galaxies (including most of the splashbacks and about a third of the non-splashbacks) have lost more gas mass than they have gained in stars that formed over the same time interval. The other half of the central breakBRDs have retained more gas than they have converted into stars, or even gained a small amount of gas since $z = 0.5$. So, for the bBRDa galaxies that have lost significant amounts of gas, some of that gas may have been converted into stars, but even more gas is still lost. Overall, the upper right panel of Figure \ref{fig:z0_history} shows that the bBRDas have formed fewer stars since $z=0.5$ than galaxies in the parent sample, although the difference is smaller when comparing to parent galaxies with similar high SFR concentration.

The star formation rate evolution for the central breakBRDs (lower left of Figure \ref{fig:z0_history}) follows a trend similar to that of the total gas mass: the central breakBRD analogues experience a higher reduction of SFR compared to the parent sample. This remains high even compared to the highly-SF-concentrated parent sample. However, the amount of reduction is larger for the breakBRDs: the median value of SFR loss is $0.55$ dex, which amounts to a reduction in star formation of $72\%$, with losses up to $1.25$ dex ($94\%$ SFR reduction). It is noteworthy that excluding splashback galaxies has less of an impact on the evolution of the SFR than on the gas content: when removing splashbacks, the median + MAD is  $-0.48 \pm 0.22$ for the breakBRDs.

Studying the evolution of the central black hole is particularly interesting as AGN related feedback can strongly influence both the central regions as well as the more extended gas distributions of galaxies. In Figure \ref{fig:BHmass} we show that the breakBRD analogue galaxies fall on the black hole mass -- stellar mass relation of the parent; in Figure \ref{fig:z0_history}, we study their black hole growth since $z=0.5$ (lower middle). While the breakBRD analogues have no unusual black hole masses, their growth history from before they were breakBRD analogue galaxies could point to whether their central black hole properties are correlated with their centrally concentrated gas and SFR distributions. The black hole growth in Figure \ref{fig:z0_history} shows similar trends to the gas mass and star formation rate growth when comparing the breakBRD analogues to their parent sample in that the growth is less; however, the difference for black hole growth is much less than for the other quantities. This is reflected in the parent and analogue medians ($0.23 \pm 0.05$ vs. $0.29 \pm 0.09$, and removal of splashbacks only brings the bBRDas to $0.24 \pm 0.06$). We note that we reach similar conclusions when comparing the maximum black hole accretion rates in each galaxy's recent history (see Section \ref{sec:bhs}). Black holes are therefore unlikely to be the main driver of the breakBRD state.

Lastly, we measure the growth of the dark matter halos between $z=0.5$ and $z=0$ (lower right panel of Figure \ref{fig:z0_history}). While central galaxies in the parent sample grow in halo mass, as significant fraction (predominantly splashbacks) of breakBRD centrals have lost halo mass, while the rest of the sample has only grown slightly. For the latter, we find that the halo mass growth is approximately proportional to the stellar mass growth over this time interval. In particular, the decrease in halo mass that a subset of the central breakBRDs experience, but also the relatively small growth in halo mass, suggest that environmental effects could have played a part in forming the current state of these galaxies. For the splashback breakBRD galaxies this is very likely the case. For the non-splashback central breakBRD analogues, the environment may have played a more subtle role in their formation. However, as the current environment of the breakBRDs is very similar to that of the parent sample, this may indicate that the past environment of breakBRDs may have been different.

By following the change in galaxy properties, we see that for the central galaxy samples, both the gas mass and SFR has decreased more in the breakBRD sample than in the parent sample, and the black holes masses, stellar masses, and halo masses have grown less.  We see that these differences are partially ameliorated when restricting the parent to high SFR concentration, but that even when the breakBRD centrals are compared only to this subsample of parent centrals, differences remain (see e.g. the running medians in Figure \ref{fig:z0_history}).  The largest difference is in the change in the SFR over time, likely because breakBRDs are selected to not only have centralized SF, but also to have red disks.

Our results indicate that before reaching the breakBRD state, the $z=0$ breakBRD sample experience a process (or processes) that removes gas and reduces or even stops star formation outside the central regions between at least $z=0.5$ and the present day. When we reproduce Figure~\ref{fig:z0_history} using the growth since $z=0.1$, we find that while the trends are similar, the differences between the bBRDa and the parent samples are much smaller. Moreover, while the breakBRD galaxies have lost gas and SFR since $z=0.1$ (1.3 Gyr ago), this is much less than what they have lost since $z=0.5$ (5.2 Gyr ago). We reiterate that there is little-to-no overlap in the galaxies of the breakBRD samples at these different redshifts (see Table~\ref{tab:overlap}). We therefore conclude that the process involved must have affected breakBRD galaxies over a significant time range and is likely to have gradually changed the galaxies to the breakBRD state.

We now look more closely at the intrinsic breakBRD sample at $z=0.5$ and use that sample to study the future evolution of breakBRD galaxies.

\section{breakBRD Analogues at Higher Redshifts}\label{sec:z05}

In the previous section, we focused on the breakBRD analogue population at $z=0.0$ and looked at what these galaxies were like in the past. We have also applied the selection criteria laid out in Section \ref{sec:methods} to IllustrisTNG galaxies at $z=0.5$ ($\sim5.2$ Gyr ago) and $z=0.1$ ($\sim1.3$ Gyr ago) and $z=0.03$ ($\sim$0.48 Myr ago), giving us new parent and breakBRD analogue samples. The number of galaxies in these samples is given in the righthand columns of Table \ref{tab:pop}. The existence of analogues across more than 5 Gyrs indicates that  breakBRDs galaxies exist at a range of redshifts. Though not necessarily common, they may therefore be a part of a regular evolutionary path. 

Two possibilities for this existence at multiple snapshots, gigayears apart, immediately spring to mind: either the same galaxies exhibit breakBRD properties for a long period of time, or the breakBRD state is a (short) phase in galaxy evolution that happens for some or even most galaxies. We discuss both these possibilities when studying the future of breakBRD galaxies in Section \ref{sec:future}.

In the following sections, we briefly show the similarity of $z=0.5$ breakBRDs' properties to those at $z=0$, and then look at the future of the $z=0.5$ galaxies. We also consider the future of the $z=0.1$ sample. For brevity, we do not show the properties of the $z=0.1$ or $0.03$ bBRDa galaxies, but they follow the same patterns as the redshift 0 and 0.5 samples.

\subsection{Properties}\label{sec:z05prop}
\begin{figure}
\centering
\includegraphics[width=\columnwidth]{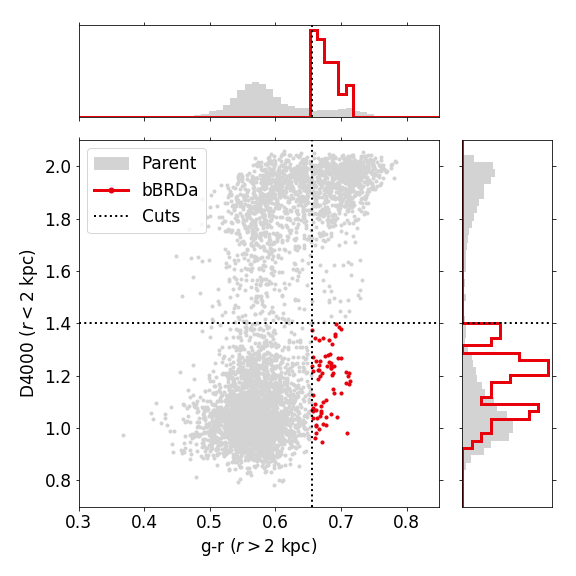}
\caption{\label{fig:z05_D4000_color} D$_n$4000 in the inner 2~kpc vs $g - r$ color in the $r > 2\mathrm{\ kpc}$ disk for our selection at $z=0.5$. The histograms at the top and right are normalized. The parent sample is shown in grey, while the $z=0.5$ breakBRD analogues are red. Dashed black lines show the $\mathrm{D}_n4000<1.4$ and $g-r > 0.655$ selection cuts. Note that the histogram bins used for plotting straddle these cut, and the axis limits are different from Figure \ref{fig:D4000_color}.}
\end{figure}
\begin{figure}
\centering
\includegraphics[width=\columnwidth]{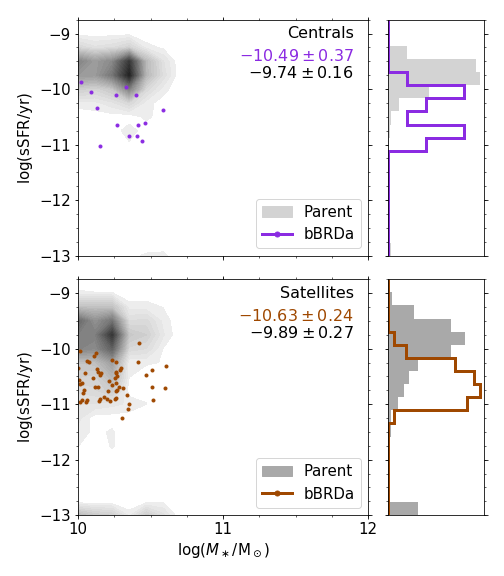}
\caption{\label{fig:z05_massSFR}Total instantaneous specific star formation rate vs total stellar mass for central galaxies (top) and satellites (bottom) in our $z=0.5$ selection. The parent sample is represented by the grey contours and the side histograms are normalized. Galaxies with no star formation were set to $\log (\mathrm{sSFR/yr})=-13$.}
\end{figure}

Both Table \ref{tab:pop} and Figure \ref{fig:z05_D4000_color} show that there are more galaxies at $z=0.5$ that have central $\mathrm{D}_n4000<1.4$, and therefore recent star formation, which is consistent with our expectations for increased star formation rates at this redshift compared to $z=0$. There are fewer galaxies at $z=0.5$ with red disks, making the breakBRD analogues at $z=0.5$ an even more unusual sample than at $z=0.0$. 
This is not unexpected, as the increased star formation will make disks tend to bluer. While we compute g-r color magnitudes in the local frame, our selection cut ($g-r>0.655$) is not adjusted for changes in galaxy population composition. These slight differences, however, do not affect the overall behavior of the selection cuts, as we also select galaxies at extreme ends of the distributions at $z=0$. The fraction of analogues that are centrals is also lower, being 19\% instead of the 37\% at $z=0.0$. This is despite a relatively consistent fraction of centrals in the parent samples at both redshifts. 

In order to compensate for correlations with stellar mass, we again weight the $z=0.5$ parent satellite and central subsamples to match the mass distribution of the corresponding breakBRD analogues, just as we did with our $z=0.0$ samples. The stellar mass distributions of both the parent and bBRDa galaxies are similar to those at $z=0.0$. Figure \ref{fig:z05_massSFR} shows the $z=0.5$ stellar mass-specific star formation rate relationship, which can be compared to Figure \ref{fig:massSFR}. Once again, galaxies with no (or particularly low) star formation were set to $\log (\mathrm{sSFR/yr})=-13$. We see the same trend here as we did in Figure \ref{fig:massSFR}, where the breakBRD analogues tend to be at lower sSFR than their weighted parent samples. For the satellites, this difference is even more pronounced at $z=0.5$---where the parent is more clustered---than at $z=0.0$.

Briefly, we find the redshift 0.5 analogues generally follow the same trends with respect to their weighted parent as the $z=0.0$ galaxies in Figures \ref{fig:massSFR}--\ref{fig:environment}. There is no significant difference between the black hole mass distributions, just like at $z=0.0$. Despite the samples being smaller, the breakBRD analogues at $z=0.5$ have dense gas masses consistent with the trends seen in Figure \ref{fig:gasmass}: the breakBRD analogues tend to have lower total gas masses than their parents, but an equivalent amount of dense gas in their central regions.

In some cases the trends seen at $z=0.0$ are more pronounced at $z=0.5$. Using the concentration measures introduced in Figure \ref{fig:multipanel}, the $z=0.5$ breakBRDs have dense gas and SFR concentration distributions that are even more clustered at the high end. 
The satellite fraction remains roughly constant for the parent sample but increases by 20\% for the breakBRDs.
Central breakBRDs at $z=0.5$ have a stronger tendency to live in high density environments than at $z=0$, where there was only a slight enhancement over the parent sample (see Figure \ref{fig:environment}). 
The primary goal of the $z=0.5$ sample is to see what happens to breakBRD analogues as they continue to evolve, and while it is meaningful that the two bBRDa populations have similar properties (see Section \ref{sec:quenching}), a detailed characterization of the redshift 0.5 sample is not relevant to this goal.

\subsection{The Future of breakBRDs}\label{sec:future}
\begin{table}
\begin{center}
\caption{Overlap in breakBRD analogue samples}\label{tab:overlap}
\begin{tabular}{lrrr}
\hline
\hline
Redshifts & $\Delta t$ (Gyr) & Number & Percentage \\
\hline
0.0 \& 0.03 & 0.48 & 135 & 46.9 \\
0.03 \& 0.1 & 0.86  & 126 & 51.0 \\
0.0 \& 0.1 & 1.3 & 63 & 25.5\\
0.1 \& 0.5 & 3.9 & 2  & 2.8 \\
0.03 \& 0.5 & 4.7  & 1 & 1.4 \\
0.0 \& 0.5 & 5.2  & 0 & 0.0 \\
\hline
\end{tabular}
\end{center}
NOTE. - Table is ordererd by the time elapsed between the two redshifts. Percentages are calculated relative to the higher redshift, using the bBRDa numbers in Table \ref{tab:pop}.
\end{table}
\begin{figure}
\centering
\includegraphics[width=\columnwidth]{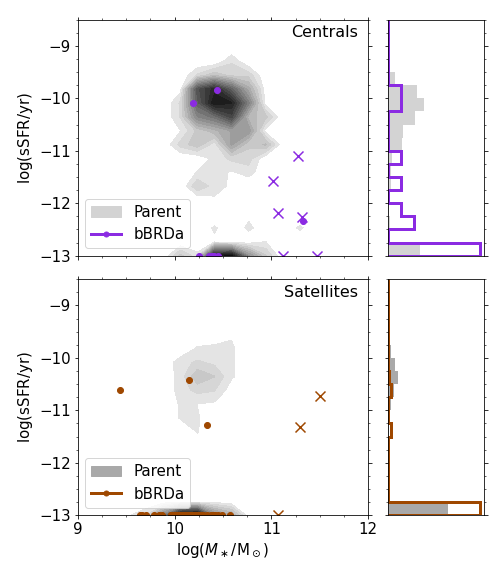}
\caption{\label{fig:massSFR-tracked}Same as Figure \ref{fig:z05_massSFR} but the parent and breakBRD samples selected at $z=0.5$ have been tracked forward to $z=0.0$. Parent galaxies maintain the weighting applied in Figure \ref{fig:z05_massSFR}; see text for more details. X's mark breakBRD galaxies that have merged with one of the parent sample by $z=0.0$.}
\end{figure}

In Table \ref{tab:overlap} we provide the overlap in bBRDa samples, i.e. the number of galaxies that are breakBRDs at two epochs, at our four redshifts: 0, 0.03, 0.1, and 0.5. 
It is out of the scope of this paper to find breakBRD analogues at every TNG snapshot in order to quantify the length of the breakBRD state in detail, but we can draw some conclusions based on our limited redshift sampling.
The overlap between the $z=0.0$ and $z=0.03$ samples shows that ${\sim}50\%$ of the breakBRDs at $z=0.03$ are still in that state 0.48 Gyr later, at redshift 0. This percentage drops to $\sim$25\% over the 1.3 Gyr span between $z=0.1$ and 0.0. This strongly suggests a relatively short-lived breakBRD state of a few hundred Myr up to ${\sim}2$ Gyr. Moreover, the galaxies that are in the breakBRD state for two of our sampled redshifts include both centrals and satellites. This suggests that there may be similar physical processes at work that facilitate the formation of a breakBRD state for both central and satellite galaxies.

We tracked our $z=0.5$ samples forward to $z=0.0$ to study the galaxies' evolution after the breakBRD state. In Figure \ref{fig:massSFR-tracked} we again plot total stellar mass and specific SFR, but for this tracked set of galaxies. Galaxies with no  or very low star formation were once again set to $\mathrm{\log(sSFR/yr)} = -13$. Parent galaxies have the same weights applied to them as assigned in Figure \ref{fig:z05_massSFR}. About 20\% of the parent galaxies have merged with another parent between $z=0.0$--$0.5$; descendent parents are only counted once, and are weighted according to the most massive of their progenitors in the $z=0.5$ parent sample. Some of the breakBRD analogues have also merged with galaxies from the parent sample; these are indicated with crosses in Figure \ref{fig:massSFR-tracked}. The bBRDa that have merged with a parent galaxy are always the less massive of the progenitor pairs. No breakBRD galaxies have merged with another breakBRD. 

In Figure \ref{fig:z05_massSFR}, all but four, or 1.7\%, of the breakBRD analogues have $\mathrm{\log(sSFR/yr)} \gtrsim -11$, which we use as our delineation between the star-forming and quiescent galaxy populations \citep[see e.g.][]{Cassata10, Tamburri14, Matthee19, Katsianis2019}. Three of these are centrals. When tracked to redshift 0 (Figure~\ref{fig:massSFR-tracked}), 85.7\% of central breakBRD analogues (12 out of 14), and 94.8\% of satellites (55 out of 58) have $\mathrm{\log(sSFR/yr)} < -11$; therefore, most of the galaxies that were breakBRDs at $z=0.5$ end up quenched. 

Determining the fraction of parent galaxies that transition from star forming to quiescence is more complicated, because of mergers. To determine if a tracked galaxy was star forming at $z=0.5$, we use the sSFR of the most massive progenitor in the sample, akin to how we handled the weighting. Again, some bBRDa galaxies have merged with redshift $z=0.5$ parent galaxies, but are always the less massive of the pair and so have no impact on whether a descendent galaxy was considered star-forming at $z=0.5$. With weighting applied, the fraction of parent centrals and satellites that quench by redshift $z=0$ is 25.3\% and 51.3\%, respectively. Unsurprisingly, two proportion z-tests indicate that the quenched fraction is significantly higher for the bBRDa galaxies at the $\alpha=0.01$ level.

Though not shown, we also track forward the redshift $z=0.1$ breakBRD analogue and perform the same analysis. At $z=0.1$, these galaxies display the same sSFR pattern as seen in Figures \ref{fig:massSFR} and \ref{fig:z05_massSFR}; yet when tracked to $z=0.0$, these galaxies have quenched to a lesser extent. Of the satellites, 51.1\% of the bBRDa galaxies have $\mathrm{\log(sSFR/yr)} < = -11$, compared to 10.7\% of the weighted parent. For the centrals, 26.0\% of the bBRDa have $\mathrm{\log(sSFR/yr)} < = -11$ versus 4.5\% of the parent. Both fractions are significantly different at $\alpha=0.01$. Preferential quenching in satellites is not unexpected, because in satellites, the breakBRD state may indicate strong ram pressure stripping or tidal effects which could rapidly quench the satellite. Such a straightforward explanation would not apply to the centrals, however.

We see a similar set of properties for the breakBRDs at redshifts 0, 0.1, and 0.5, which suggests that the breakBRD selection criteria chooses a consistent set of galaxies. We therefore feel confident that the quenching experienced by the $z=0.5$ breakBRDs implies that all low redshift galaxies in the breakBRD state will eventually quench.

\section{Discussion}\label{sec:discussion}

The following discussion will use the results put forth in Sections \ref{sec:z00} and \ref{sec:z05} to answer a series of questions about the formation and evolution of breakBRD analogues.  Finally, we will briefly compare the simulated analogues to the observational sample.

\subsection{Central Enhancement vs Outer Deficit }\label{sec:conc-vs-deficit}

In Figure \ref{fig:multipanel} and the accompanying discussion, we find that the breakBRD analogues have SFR and gas distributions that are centrally concentrated. In the case of SFR and dense, star formation-eligible gas, these concentrations can be quite high, with at least 40\% of each contained within the central 2 kpc-radius.
Our selection criteria were designed to choose galaxies with highly concentrated star formation, and these stars must be forming out of dense gas, so high concentrations are expected. Note, however, that the parent sample also includes non-breakBRD galaxies with highly concentrated SFR and dense gas: the bBRDa galaxies differ from these because these non-breakBRD SFR-concentrated galaxies also have some star formation in their outer regions. We now ask: did a process cause breakBRD galaxies to have higher central concentrations? Or do they appear this way because gas was removed from their outskirts?

Figure \ref{fig:gasmass} is useful for distinguishing central gas enhancement from outer gas deficit, as the dense gas ratio from Figure \ref{fig:multipanel} is separated into its components. Here we find that the breakBRD analogues have roughly the same gas mass in their interiors as parent galaxies weighted for the same stellar mass, despite being lower in dense gas mass overall. We interpret this as a strong indication that gas has been removed from the outskirts ($r > 2$~kpc) of the breakBRD galaxies. 

It could also be the case that gas has been funneled from the outskirts into the centers of the breakBRD analogues, rather than stripped from the galaxies completely. 
Though not shown here, we examined the mass of total gas in a manner similar to Figure \ref{fig:gasmass} and found a similar trend: bBRDa galaxies have a lower total gas mass, but inner gas mass equivalent to the parent. This structure, as well as the patterns in Figure \ref{fig:multipanel} could also be explained by gas redistribution.
However, Figure \ref{fig:z0_history} shows that breakBRD analogues tend to have lost more gas than is accounted for by their growth in stellar mass. It is therefore extremely likely that the breakBRD galaxies have lost significant amounts of gas from their outskirts; gas which therefore does not get converted into stars or accreted onto the central black hole. Because of the reduced amount of gas in the outskirts and normal central gas mass, breakBRD analogues appear centrally concentrated.

\subsection{The Formation of breakBRDs}\label{sec:formation}

The selection criteria laid out in Section \ref{sec:bbrda} certainly select a unique sample of galaxies, and it is instructive to investigate how these galaxies form. Indeed, the stripping suggested by Figure \ref{fig:gasmass} (and discussed in Section \ref{sec:conc-vs-deficit}) motivated Section \ref{sec:properties_drivers}, as black holes, environment, and merger history could all possibly explain a deficiency of outer gas (see Section \ref{sec:intro}).
While we discuss these processes separately, here we endeavor to synthesize what the results of \ref{sec:properties_drivers} tell us about the formation of breakBRD analogues.

First, we consider black holes. With Figure \ref{fig:BHmass}, we find no difference between the black hole mass of $z=0$ bBRDa and parent galaxies. Their strong agreement persists when we remove the power law dependency on stellar mass, and also holds for breakBRD analogues at higher redshift. We also looked at black hole growth over the past 5 Gyr with Figure \ref{fig:z0_history}. We do not see a strong difference between parent and bBRDa galaxies in terms of their black hole growth: though the bBRDa centrals tend to have black holes that grow more slowly, the apparent difference is not significant. 

Second, we examined the fraction of galaxies that experienced at least one merger of a given size, including very small mergers that could be instead defined as clumpy cosmological accretion. We only find a statistically significant difference for the fraction of centrals that have had at least one minor accretion event (mass ratio of $1/100$--$1/10$): bBRDa centrals are less likely to have had such an accretion event than parent centrals. This leads us to conclude that the highly centrally concentrated SFR exhibited by breakBRD galaxies is unlikely to be caused by mergers and other growth history. On the other hand, this low frequency could be related to a general decrease in cosmic gas accretion (including smooth accretion) which may then have induced the low total gas mass discussed above, and the low halo and stellar mass growth for breakBRD galaxies in Figure \ref{fig:z0_history}. This effect would still not fully explain the breakBRD state, however, given that most breakBRD galaxies have lost significant amounts of gas.

Finally, satellite galaxies may have outlying gas stripped by environmental processes (either gravitational or hydrodynamical).  We do see a much higher satellite fraction and splashback fraction in the breakBRD analogue sample than the parent sample, indicating that environmental processes could be an important process forming breakBRD galaxies. Also, when we examined the change in gas mass of central $z=0$ breakBRD galaxies before they reached the breakBRD state (Figure~\ref{fig:z0_history}), we found that the outer gas mass had generally decreased from $z=0.5$.  They have also experienced a commensurate change in dark matter mass, possibly hinting at subtle environmental effects forming even the central breakBRD population.

However, when measuring the density of their present day environment (Figure \ref{fig:environment}), the median + MAD values overlap for the
non-splashback breakBRD analogues and the non-splashback parent sample.  
Though not shown, we also quantified their environment through the distance to the nearest more massive neighbor and to the nearest cluster. Here we also found significant overlap between the median + MAD of the central breakBRD and parent samples. Therefore, while environmental processes can explain the properties of the satellites and splashback galaxies in the breakBRD sample, 
if the environment has an important role in the formation of central breakBRD analogues, this may indicate that the environment of breakBRD central galaxies has changed. 

The strongest difference between breakBRD analogues and their TNG parent sample seems to be environment; however, the difference is not a ``smoking gun.'' Importantly, we stress that throughout our findings, breakBRD analogues are more like each other (across the satellite-central divide) than galaxies in the parent sample.  For example, the median dense gas mass and concentration of breakBRD satellites and centrals are similar, and both are significantly larger than the concentrations of the parent subsamples. While environment may be an influencial factor in the formation of breakBRDs, the exact way in which it exerts this influence appears to be subtle and worthy of further, more detailed investigation beyond the gross statistics explored herein.

\subsection{Quenching vs Stochastic Star Formation}\label{sec:quenching}
\begin{figure}
\includegraphics[width=0.5\textwidth]{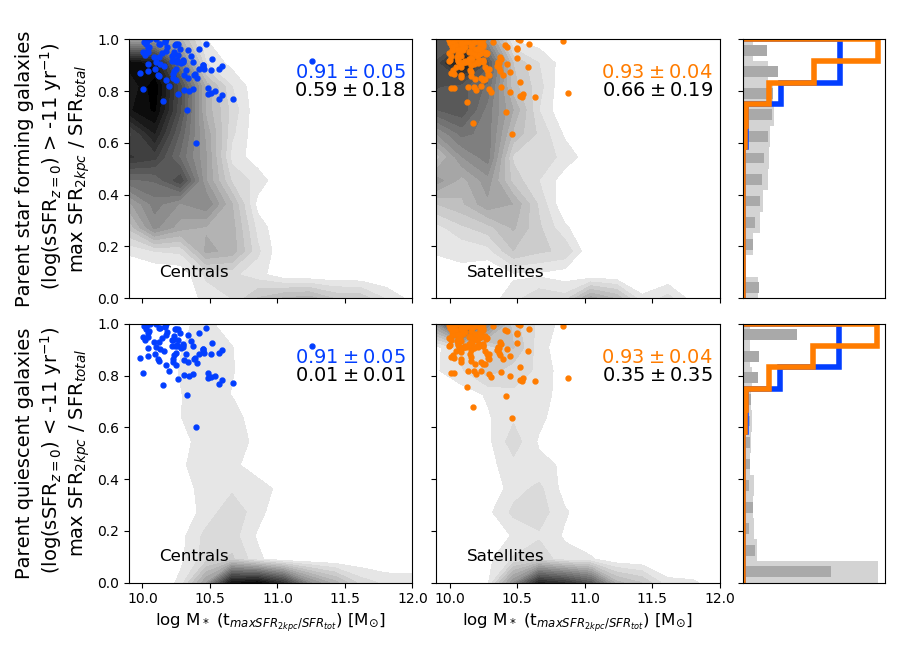}
\caption{\label{fig:traceback1}Maximum star formation rate concentration, max(SFR$_{\rm 2 kpc}/$SFR$_{\rm total}$), of each galaxy, either since $z=0.5$ for star-forming parent galaxies (top), or since having log(sSFR/yr) $ \geq -10$ for quenched parent galaxies (bottom) against the stellar mass they had at the time of peak SFR concentration. The plot is broken into centrals (left) and satellites (right). Overplotted is the $z=0$ breakBRD analogue sample, also shown with their historically highest SFR concentration and corresponding stellar mass since $z=0.5$. In contrast to other figures, the breakBRD sample is not included in either the star-forming or quiescent parent sample. Note that in contrast to most other figures, here we show \emph{all parent galaxies without weighting}.}
\end{figure}
\begin{figure}
\includegraphics[width=0.5\textwidth]{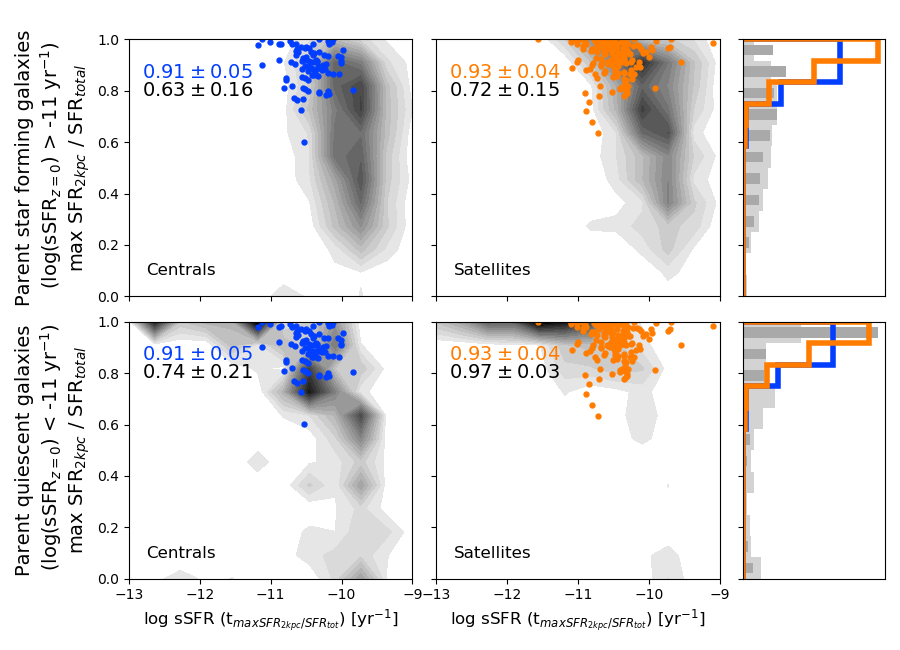}
\caption{\label{fig:traceback2} Similar to Figure \ref{fig:traceback1} but with peak SFR concentration plotted against the sSFR at the time of peak SFR. Moreover, here we show the parent galaxy sample distribution \emph{weighted} to have the same present-day stellar mass distribution as the $z=0.0$ breakBRD analogue sample. Star-forming/quenched satellites/centrals are all weighted separately. As in Figure \ref{fig:traceback1} the breakBRD sample is not included in either the star-forming or quiescent parent sample.}
\end{figure}

One benefit of cosmological simulations is that we can answer the question ``what will happen to this galaxy in the future?''. This is exactly the question we asked of the $z=0.5$ breakBRD analogues in Figure \ref{fig:massSFR-tracked}, where we concluded that breakBRDs largely became quenched by $z=0.0$. Though we conclude that the $z=0.5$ breakBRDs quench, we have not determined whether the breakBRD state is common to galaxies in TNG100.

We now consider whether or not the breakBRD state is a common state in the lives of galaxies, either generally, only within the mass range typical to the bBRDa sample (see Figure \ref{fig:masshistogram}), or specifically for galaxies that will quench. 
Because the breakBRD state is phenomenologically characterized by high SFR concentration (Figure \ref{fig:multipanel}),
we can asses whether or not a galaxy \textit{could} have ever been in the breakBRD state by looking at maximum SFR concentration each galaxy has reached during its history.  
We first split our $z=0.0$ parent sample into star-forming and quiescent subsamples using $\log(\mathrm{sSFR/yr}) = -11$ as the boundary. For the star-forming sample, the maximum SFR concentration of the parent sample is determined across \textit{all} IllustrisTNG outputs between $z=0.5$ and $z=0$. Redshift 0.5 is chosen as the endpoint because there is no overlap in the breakBRD population between $z=0.5$ and $z=0.0$.  For the quiescent sample, we find the maximum SFR concentration since the galaxy was on the star-forming main sequence (at $\log(\mathrm{sSFR/yr}) > -10$), which can occur more recently than $z=0.5$ or at higher redshift.
We reiterate that this search is performed at all outputs, not just those redshifts at which we found breakBRD analogues. We note that high SFR concentration does not necessarily mean a galaxy \textit{would} meet the breakBRD selection criteria, as this additionally requires red disk color (see Section~\ref{sec:bbrda}). 

Figure \ref{fig:traceback1} shows the maximum SFR concentration for the \textit{unweighted} parent sample in grey, while Figure \ref{fig:traceback2} shows maximum SFR concentration for the \textit{weighted} parent sample. Note that the star-forming/quenched satellites/centrals have their weights calculated separately, making their weightings slightly different than those used in Section \ref{sec:z00}. Additionally, the parent sample excludes the breakBRD subsample in both of these figures, in contrast to the rest of this work. The central (blue; left) and satellite (orange; right) $z=0.0$ breakBRD analogues are repeated in both Figures \ref{fig:traceback1} and \ref{fig:traceback2} as a reference, with their peak SFR since $z=0.5$.

In Figure \ref{fig:traceback1} we show, for the unweighted parent and bBRDa samples, the maximum SFR concentration against the total stellar mass at the time of maximum SFR concentration. 
Removing the mass weighting allows us to explicitly see the distribution of SFR concentration over our full stellar mass range, for both star-forming and quiescent galaxies, and compare these to our breakBRD samples.

For the weighted parent sample shown in Figure \ref{fig:traceback2} we plot the maximum SFR concentration against the \textit{total sSFR} at that time.  This is because in addition to SFR concentration, breakBRD galaxies tend to cluster in total sSFR space (Figures \ref{fig:massSFR} \& \ref{fig:z05_massSFR}).  The combination of star-forming center and red disks tends to make breakBRDs sit somewhat below the star-forming main sequence while still well above the sSFR of quenched galaxies, so this may give us more discriminating power in our comparison with the parent sample.

We first focus on the star-forming galaxies in the top panels of Figures \ref{fig:traceback1} and \ref{fig:traceback2}.  Whether we consider the weighted (Figure \ref{fig:traceback1}) or unweighted (Figures \ref{fig:traceback2}) samples, we see that the maximum concentration of star-forming galaxies extends to much lower values than the breakBRD sample.  This is also reflected in their median + MAD values.  The central and satellite breakBRD MAD ranges overlap with each other, but generally not with their respective parent populations. The breakBRDs only have overlapping median+MADs with the weighted quenched parent population.  We note that at the time of maximum concentration, the sSFR of star-forming parent galaxies extends to higher values than the breakBRD analogue sample. This is very dramatic in the central parent sample, but can be seen even in the star-forming satellite parent galaxies. We therefore conclude that the breakBRD state is unlikely to be a result of stochastically-distributed star formation in galaxies on the star-forming main sequence.  This comes as no surprise since we found that breakBRD galaxies (identified at $z=0.5$) are likely to quench.

We next focus on the quenched galaxies in the bottom panels of Figures \ref{fig:traceback1} and \ref{fig:traceback2}.  When we examine the unweighted sample, we clearly see that most quenched galaxies have not passed through a breakBRD state: there is a the large number of galaxies with low concentrations, especially at higher masses.  Focusing on the unweighted parent sample in Figure \ref{fig:traceback1} allows us to highlight the unusual mass distribution of breakBRD galaxies with respect to quenched galaxies---they tend to be lower mass. At the mass range typical for breakBRD galaxies, there seems to be a high fraction of quenched galaxies that have experienced centrally concentrated star formation in their past.

When looking at the weighted sample in Figure \ref{fig:traceback2}, which emphasises the breakBRD mass range, we see that nearly all quenched satellites and about half of the quenched centrals have had high SFR concentrations, indicative of a breakBRD-like state. Looking at central galaxies in detail, we see that the median is below most of the range of the breakBRD sample, but because the distribution is so extended there is significant overlap in the concentrations. This state is therefore more common among quenching galaxies with lower stellar masses, but even among low-mass quenching galaxies the breakBRD state is far from ubiquitous.

Because both central star formation (measured using D$_n$4000) and a red disk is required for the breakBRD state, by only looking at SFR concentration, we are being conservative in our estimate of the historic overlap in the breakBRD and parent populations.  BreakBRD galaxies are likely more unique than the median + MAD values in Figure \ref{fig:traceback2} indicate.  We can begin to see this when we examine the range of sSFR values in the parent sample in Figure \ref{fig:traceback2}.  We see that for galaxies quenched at $z=0$, the sSFR at maximum SFR concentration can be much lower than that found in the breakBRD sample, indicating that these galaxies might not meet the the star-formation threshold for D$_n4000 < 1.4$.

This brings us to an important caveat in our comparison.  It is outside of the scope of this work to search for every breakBRD galaxy at all of the IllustrisTNG outputs.  All we are finding for the parent sample is the simulation snapshot with the maximum SFR concentration.  There may be other times when the SFR concentration is still quite high and the total sSFR is more in line with that found for breakBRDs.  We can only draw conclusions about the fraction of galaxies that \textit{could} reach the breakBRD state criteria based on this simple proxy.

By searching for the maximum SFR concentration of quenched parent galaxies, we conclude that the majority of these galaxies do \textit{not} experience a breakBRD state. Moreover, when we look at quenched parent galaxies at similar stellar masses as the breakBRDs, at least half of central galaxies never reach high SFR concentrations.This difference in SFR concentrations between high- and low-mass quiescent galaxies does suggest that these galaxies can experience different quenching pathways in IllustrisTNG. Nevertheless, we conclude that breakBRD galaxies remain an unusual sample of quenching galaxies, even when attempting to align our samples to the same point in their evolution. 

\subsection{Comparison with the Observational Sample}\label{sec:obsv_compare}
In this paper we have specifically chosen to compare galaxies within the TNG100 simulation, rather than compare simulated galaxies directly to observed galaxies.  This way, we  mitigate any ``systematic'' errors in the simulation, which we know does not exactly reproduce the observed universe. For instance, for the stellar mass range of the breakBRDs, the TNG black hole mass-stellar mass relation tends towards the higher end of what is observed \citep{Habouzit2020}, and the satellite quenched fraction is higher than seen in SDSS \citep{Donnari2020}.  That said, because we are using this work to draw conclusions about the evolution of breakBRD galaxies, it is worthwhile to make some comparisons between the simulated and observed breakBRD galaxy samples.

Looking at the stellar mass distribution of the breakBRD observed and analogue samples, we find that they are quite similar. Although we require that IllustrisTNG galaxies have M$_*$ > 10$^{10}$ M$_{\odot}$, about 80\% of observed breakBRDs are above this mass. The majority of observed breakBRDs are between 10$^{10}$ - 10$^{10.5}$ M$_{\odot}$ (see Figure 3 in TT20), roughly agreeing with the tendency of breakBRD analogues to have lower masses within our selected range of 10$^{10}$ - 10$^{12}$ M$_{\odot}$ (Figure \ref{fig:masshistogram}).

\begin{figure}
\centering
\includegraphics[width=\columnwidth]{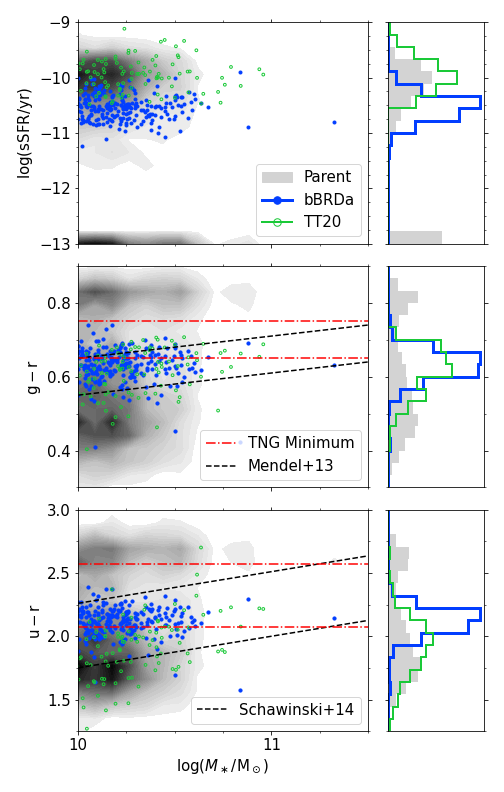}
\caption{\label{fig:color-compare}Whole-galaxy sSFR (top), $g-r$ (middle) and $u-r$ (bottom) for the full breakBRD analogue sample (blue solid) and its weighted parent (greyscale). The sSFRs and colors of the observed breakBRD sample \citep{tuttleBreakBRDGalaxiesGlobal2020} with $\log(M/\mathrm{M_\odot}) > 10$ are shown as green open circles and thin lines. In the lower two plots, black dashed lines show the ``green valley'' \citep{mendelPhysicalPictureStar2013, schawinskiGreenValleyRed2014}, and red dot-dashed lines bracket the ``valley'' present in the colors of the un-weighted TNG parent; see text for more details. }
\end{figure}

Figure \ref{fig:color-compare} compares the sSFR and color distributions of the observed and simulated breakBRD samples, as well as the weighted TNG parent sample. Note that a mass cut of $\log(M_\ast) > 10\ \mathrm{M_\odot}$ has been applied the observed breakBRD sample, in agreement with the mass cut used to select the TNG parent sample (Section \ref{sec:parent}).

We first focus on the top panel, which compares the sSFR of the aforementioned samples. The TNG data combines satellites and centrals from Figure \ref{fig:massSFR}. The observed parent is omitted, but its star-forming component (log(sSFR/yr)$ > -10$) has an sSFR distribution that is very similar to the weighted TNG parent's. With this in mind, we can see that the sSFR distribution of observed breakBRDs (green points) is similar to that of its star-forming parent sample, while the simulated breakBRDs (blue points) have sSFRs lower than the median of their parent samples.  This may indicate that the breakBRDs in the simulation have already begun quenching while the observed galaxies are still on the star-forming sequence.  However, the global sSFR from \citet{brinchmannPhysicalPropertiesStarForming2004} is used in TT20.  This uses the emission lines in the SDSS spectra from the central region (the region covered by the spectral fiber) as well as the optical colors from the SDSS photometry inside and outside the fiber regions to place galaxies on a global sSFR relation. \citet{brinchmannPhysicalPropertiesStarForming2004} note however, that the likelyhoods of SFR/L$_i$ for redder colors are broader and sometimes multipeaked, and the SFR/L$_i$ is therefore less well constrained. Because the breakBRD galaxies have unusual central star formation rates for their red disk colors, the global sSFRs from \citet{brinchmannPhysicalPropertiesStarForming2004} may be more uncertain for these galaxies.

In the lower panels of Figure \ref{fig:color-compare}, we compare the global $(g - r)$ and $(u - r)$ colors of the simulated and observed galaxies.  We see that particularly in the $(g - r)$ galaxy color, the observed (green points) and simulated (blue points) breakBRD galaxies are quite similar, although more of the observed breakBRDs extend to bluer colors.  In addition we have shown the observed ``green valley" with black dashed lines, as fit using color-stellar mass diagrams \citep{mendelPhysicalPictureStar2013, schawinskiGreenValleyRed2014}.  We also have found the minimum of the color distributions in the TNG100 parent sample as a function of stellar mass.  It is nearly constant with mass, and we denote the $\pm$0.5 dex region with red dashed lines to show a rough "green valley" for the simulated parent sample.  We see that the simulated valley is at redder colors than the observed valley for most of the galaxies in the sample. Thus the $(g - r)$ colors for the simulated sample are bluer with respect to the parent than the observed sample. The $(u - r)$ colors are similar with respect to the valley in both the simulated and observed samples.  

Finally, we compare the environmental measures of the observed and simulated breakBRD galaxies.  The satellite fraction of observed breakBRD galaxies seems to be quite similar to the observed parent sample (40\% and 39\%, respectively), while the satellite fraction of breakBRD analogues is much larger than that of the parent sample (see Table \ref{tab:pop}).
However, for both the observed and analogue breakBRD samples, the environmental density is similar to that of their parent samples, even when including splashback galaxies (which cannot be removed from the observed sample).

In summary, we find that there is generally good agreement between the observed and simulated samples.  However, when compared to the parent samples, the sSFRs of the simulated breakBRDs tend to be lower than the sSFRs of the observed breakBRDs, and the satellite fraction of breakBRD galaxies is comparatively higher in the simulations than in the observations.

\section{Conclusion}\label{sec:conclusion}

\citet{tuttleBreakBRDGalaxiesGlobal2020} discovered an unusual sample of galaxies in observations, called breakBRDs, with red disks and recent star formation in their centers. By generating synthetic observations, we applied similar cuts to galaxies within IllustrisTNG to find galaxies in a state analogous to the observed breakBRDs. These cuts were applied at multiple redshifts: 0.0, 0.03, 0.1, and 0.5.

\begin{itemize}
    \item IllustrisTNG contains galaxies 
    that are analogous to the observational sample of TT20 in terms of their disk colors and D$_n$4000 measures (Figure \ref{fig:D4000_color}). 
    
    \item Our color- and D$_n$4000-based selection criteria find TNG galaxies with centrally-concentrated star formation (Figure \ref{fig:multipanel}), as expected from TT20.  Simulated breakBRD galaxies tend to have lower stellar masses than the parent TNG sample (Figure \ref{fig:masshistogram}).  When we weight the parent sample by its stellar mass distribution, we find that the dense ($> 0.1\ \mathrm{cm^{-3}}$) gas mass is low and centrally concentrated in bBRDa galaxies (Figure \ref{fig:gasmass})
    
    \item The bBRDa populations have a higher satellite fraction than in the parent sample (Section \ref{sec:environ}; Table \ref{tab:pop}). 
    This aligns with our idea that environmental effects could drive centrally-concentrated star formation.  We also find a somewhat higher splashback fraction in the central galaxies of the bBRDa sample compared to the central parent sample (Section \ref{sec:environ}; Table \ref{tab:splashbacks}). Together this suggests that environment may have been a driver in forming a significant fraction of our breakBRD sample.
    
    \item Central breakBRD galaxies at $z=0$ have lost significant amounts of gas, mainly from their outskirts, and an even higher fraction of star formation since $z=0.5$ (Figure \ref{fig:z0_history}). They also generally have grown less in dark matter mass, stellar mass, and black hole mass, compared to the parent sample, although the differences between the samples are not significant.
    
    \item When we only consider central galaxies that have not been splashbacks since $z=0.5$, we find no clear cause for the breakBRD state.  The black hole mass distribution and local environment are similar between the bBRDa and parent samples (Figures \ref{fig:BHmass} \& \ref{fig:environment}). Mergers are similarly prevalent in both samples, with the exception that central breakBRD galaxies have experienced less very minor mergers or clumpy cosmological accretion (Section \ref{sec:merger}). 
    We infer that any environmental driver must therefore be subtle, and possibly act over long timescales.

    \item Within IllustrisTNG, the breakBRD-analogue state is a transient one, lasting between a few hundred Myr to $\sim2$ Gyr (Table \ref{tab:overlap} and Section \ref{sec:future}). 
    
    \item We find that breakBRD analogues at $z=0.5$ largely quench by $z=0$ (Figure \ref{fig:massSFR-tracked}). Because breakBRD galaxies at multiple redshifts have similar properties, this suggests that the majority of breakBRD galaxies are in the process of quenching, and that the observed breakBRD galaxies may also be quenching.
    
    \item We argue that breakBRD galaxies are a unique population of quenching galaxies. They are of lower stellar mass than most quenching or quiescent galaxies. By-and-large, both star-forming and quiescent galaxies do not reach the same high SFR and dense-gas concentrations exhibited by the breakBRD samples (Figures \ref{fig:traceback1} and \ref{fig:traceback2}) and therefore never experience a breakBRD state. However, present-day low-mass quiescent galaxies are more likely to have experienced a centrally-concentrated phase in their past. 
    
\end{itemize}

This paper was motivated by the observational breakBRD sample from TT20, in the hopes that IllustrisTNG could provide an explanation for the appearance of the observed sample.  Our findings place breakBRDs in a unique space outside of the general picture of galaxy inside-out formation and quenching. Most observations, as discussed in the introduction, find that as the global star formation rate decreases, central star formation is suppressed.  We have found here that, as in observations, the fraction of these highly-concentrated breakBRD galaxies is small at any given time in TNG.  We briefly note that in Illustris and EAGLE the quenching populations are mainly composed of galaxies that have centrally concentrated star formation \citep{Starkenburg2019}, unlike TNG. Importantly, upon closer inspection, we found that\textit{a significant fraction of quenching, low stellar mass central galaxies} undergo a phase of centrally-concentrated star formation (Figure \ref{fig:traceback1} and \ref{fig:traceback2}). Therefore, understanding the breakBRD state in more detail may be an important step in understanding the process by which low mass central galaxies quench. 

It is worth noting that breakBRD galaxies are not alone in indicating that galaxies may follow several paths to red, quenched, early-type galaxies.  For example, \citet{suh2010}  identify  early-type galaxies in the Sloan Digital Sky Survey DR6, and  find  that  about  30\%  of  their  sample  are ``blue-cored''.  They find  that  these  early-type galaxies tend to be lower mass and posit that they may be formed via mergers with gas-rich galaxies, while the majority of early-type galaxies are formed via equal-mass ``dry'' mergers. \citet{evans2018} find a population of ``red misfits'' characterised by red optical colors and high sSFRs, and conclude that they are likely to be gradually quenching via internal processes.

Another interesting class of galaxies are passive spirals, first found in clusters \citep{vandenBergh1976, poggianti1999}.  More recent studies have found passive spirals across environments \citep{masters2010, bamford2009}. \citet{Bundy2010} argue that up to 60\% of spirals may pass through this phase on the way to the red sequence.  Interestingly, the authors find that passive spirals are bulge-dominated at all masses, indicating that the simple fading of disks is not a viable formation mechanism.  Given that we find that nearly half of quenching central galaxies (and most quenching satellite galaxies) in IllustrisTNG pass through a breakBRD-like state (Figure \ref{fig:traceback2}), more study of a possible connection between these galaxy classes may be warranted.

Through our analysis of breakBRD galaxies we have made testable predictions that require more detailed observations of the gas distribution of observed breakBRD galaxies.  Specifically, we argue that the observed breakBRD population will show normal central gas masses and apparent outer deficits (see Section \ref{sec:conc-vs-deficit}). Given that breakBRD analogues within IllustrisTNG appear to quench, it will also be useful to search the observed sample for additional signs of quenching.  Both of our predictions can be evaluated by searching for the gas supply of breakBRD galaxies, which should be low overall but normal in the galaxy center.

\section{Acknowledgements}
The authors would like to thank the referee for comments that improved this work. It is a pleasure to thank Shy~Genel, Vicente Rodriguez-Gomez, and Brian W.~O'Shea for valuable suggestions and discussion. The authors would like to thank the IllustrisTNG collaboration for making their data public. CK is supported by the Department of Energy Computational Science Graduate Fellowship program (DE-FG02-97ER25308). This research was supported in part through the computational resources and staff contributions provided by the Quest high performance computing facility at Northwestern University, which is jointly supported by the Office of the Provost, the Office for Research, and Northwestern University Information Technology, and by the Institute for Cyber-Enabled Research at Michigan State University. The data used in this work were, in part, hosted on facilities supported by the Scientific Computing Core at the Flatiron Institute, a division of the Simons Foundation, and the analysis was largely done using those facilities. This work was initiated as a project for the Kavli Summer Program in Astrophysics held at the Center for Computational Astrophysics of the Flatiron Institute in 2018. The program was co-funded by the Kavli Foundation and the Simons Foundation. We thank them for their generous support. The Flatiron Institute is supported by the Simons Foundation.

\software{Astropy \citep{astropycollaborationAstropyProjectBuilding2018}, FSPS \citep{conroyPropagationUncertaintiesStellar2009, conroyPropagationUncertaintiesStellar2010}, python-fsps \citep{danforeman-mackeyPythonfspsPythonBindings2014}, and SciPy (esp. NumPy \& Matplotlib) \citep{scipy2020}.}

\bibliography{breakBRD_analogues}

\end{document}